\documentclass[12pt,a4paper]{article}
\usepackage{jheppub}
\usepackage{dsfont}
\usepackage{amsmath, amssymb, cool}
\usepackage{graphicx, caption, subcaption}
\usepackage{bbm}
\usepackage{epsfig}
\usepackage{epstopdf}
\usepackage{latexsym}
\usepackage{booktabs}
\usepackage[dvipsnames]{xcolor}
\pdfoutput=1
\makeatletter
\def\@fpheader{\relax}
\makeatother

\def\F{{\cal F}}

\def\F{{\cal F}}

\def\be{\begin{equation}}
\def\ee{\end{equation}}
\def\bea{\begin{eqnarray}}
\def\eea{\end{eqnarray}}
\def\ba{\begin{array}}
\def\ea{\end{array}}
\def\ben{\begin{enumerate}}
\def\een{\end{enumerate}}

\newcommand{\dsl}{\pa \kern-0.5em /}

\def\T{\Theta}

\def\F{\Phi}

\def\pa {\partial}

\def\ms{\mathcal{S}}
\def\ma{\mathcal{A}}
\title{HEE and HSC for flavors: Perturbative structure in open string geometries}
\author{Avik Banerjee$^{a,c}$, Aranya Bhattacharya$^{b,c}$, Sabyasachi Maulik$^{b,c}$}
\affiliation[a]{Harish-Chandra Research Institute, \\
Chhatnag Road, Jhusi, Allahabad 211019, India. }
\affiliation[b]{Theory Division, Saha Institute of Nuclear Physics, \\
1/AF, Bidhannagar, Kolkata 700064, India. }
\affiliation[c]{Homi Bhabha National Institute,\\
Training School Complex, Anushakti Nagar, Mumbai 400085, India} 
\emailAdd{avikbanerjee[at] hri.res.in} 
\emailAdd{aranya.bhattacharya[at] saha.ac.in} 
\emailAdd{sabyasachi.maulik[at] saha.ac.in}

\abstract{Introduction of electric field in the D-brane worldvolume induces a horizon in the open string geometry perceived by the brane fluctuations.  We study the holographic entanglement entropy (HEE) and subregion complexity (HSC) in these asymptotically AdS geometries in three, four and five dimensions aiming to capture these quantities in the flavor sector introduced by the D-branes. Both the strip and spherical subregions have been considered.   We show that the Bekenstein-Hawking entropy associated with the open string horizon, which earlier failed to reproduce the thermal entropy in the boundary, now precisely matches with the entanglement entropy at high temperatures.  We  check the validity of  embedding function theorem while computing the HEE and attempt to reproduce the first law of entanglement thermodynamics, at least at leading order.  On the basis of obtained results, we also reflect upon consequences of applying Ryu-Takayanagi proposal on these non-Einstein geometries. }

\begin{document}

\maketitle
\flushbottom

\section{Introduction} \label{sec1}
The advent of gauge/string duality\cite{Maldacena:1997re, Aharony:1999ti} has triggered an active involvement in holographic computation of field theoretic quantities over the past couple of decades. Of particular interest is to measure quantum information in such strongly coupled gauge theories, initiated by the seminal work of Ryu and Takayanagi \cite{Ryu:2006bv, Ryu:2006ef}. The quest for the gravity counterpart of these various quantum information theoretic measures \cite{ Witten:2018zva} has indeed been an active area of research and the dictionary is far from completion. Entanglement entropy (EE) \cite{Bombelli:1986rw, Srednicki:1993im, Holzhey:1994we, Calabrese:2004eu, Calabrese:2005zw,Calabrese:2009qy, Nishioka:2009un, Takayanagi:2012kg, Witten:2018lha}, the fidelity susceptibility or Fisher information metric 
\cite{ Hayashi, Petz, Lashkari:2015hha, MIyaji:2015mia, Alishahiha:2017cuk, Banerjee:2017qti}, the Bures metric
\cite{PhysRevLett.72.3439} are few of such quantities that have been looked upon time and again.

The celebrated work of Ryu and Takayanagi, is the first instance of geometrization of the  field theoretic notion of entanglement in spacetimes with constant negative curvature and expectedly has opened up a streamline of works along this direction. It is well-known that the EE is a good measure of the amount of quantum information in a bipartite system. One way to quantify the EE is to calculate the von Neumann entropy of a system divided into two parts, namely $A$ and $B$. The von Neumann entropy of part $A$ is then defined as $S_A = - {\rm Tr}(\rho_A \log \rho_A)$,
where $\rho_A = {\rm Tr}_B (\rho_{\rm tot})$ is the reduced density matrix on $A$, obtained by tracing out system $B$ from the density matrix of the 
entire system $\rho_{\rm tot}$. The holographic prescription to measure this quantity in the bulk is given by the famous Ryu-Takayanagi (RT) formula \cite{Ryu:2006bv, Ryu:2006ef}
\be\label{HEE}
S_{A} = \frac{{ \mathcal{A}}(\gamma_A^{\rm min})}{4G_N} ~,
\ee
where $\gamma_A^{\rm min}$ is the $d$-dimensional (co-dimension 2) minimal area surface in $AdS_{d+2}$  whose boundary matches with that of the subsystem $A$ in the boundary field theory,
i.e., $\partial\gamma_A^{\rm min} = \partial A_{bdy}$ and $G_N$ is the Newton's constant in $(d+2)$ dimensions.  Eventually the domain of application of this prescription has been successfully extended to cases of arbitrary dimensions, nonstatic situations \cite{Hubeny:2007xt, Lewkowycz:2013nqa, Faulkner:2013ana, Engelhardt_2015} and asymptotically AdS spacetimes \cite{Bhattacharya:2012mi, Allahbakhshi:2013rda, Mishra:2015cpa, Bhattacharya:2017gzt}. 
For asymptotic cases, extra finite contributions appear in  EE  and have  been studied in detail. These extra terms obey relations which are surprisingly analogous
to the standard thermodynamic relations, hence going by the name of entanglement thermodynamics \cite{Bhattacharya:2012mi, Allahbakhshi:2013rda, Pang:2013lpa, Mishra:2015cpa, Bhattacharya:2017gzt, Chakraborty:2014lfa, PhysRevD.100.126004}.

In this article we look to extrapolate the dictionary to another class of asymptotically $AdS$ spacetimes --- the  \textit{open string geometries}. These geometries were first encountered in \cite{Seiberg:1999vs} in the context of string theory in non-commutative background. In our case, these geometries emerge kinematically on studying fluctuations on D-branes in certain $AdS$ backgrounds. Our goal in this paper is kind of two-fold -- firstly, to the best of our knowledge, these are the first instances of non-Einstein spacetimes where we look to implement the Ryu-Takayanagi prescription.  Furthermore, as we will show, these geometries typically violate one of the energy conditions depending upon the dimension. Also, the horizons that we study in this paper are engineered in a rather unconventional sense as compared to usual black hole formation process and the representative dual state in the boundary is in a non-equillibrium steady state (NESS) \cite{Kundu:2013eba,Kundu:2015qda,Banerjee:2015cvy}. It will be interesting to see whether the imprints of this peculiarities somehow show up in our results. Secondly, the bulk physics in open string geometries is related to the flavor physics in the dual gauge theory introduced by the insertion of D-branes. In that sense, our efforts look to extend the Ryu-Takayanagi prescription to study entanglement in flavor sector of the gauge theories. It is worth noting at this point that our work is similar in spirit to the ones carried out in \cite{Azeyanagi:2007qj,Chang:2013mca,Kontoudi:2013rla}, however in this case, we will completely restrict ourselves to the probe approximation and carry out the entire study in open string geometries. So the background in our case just provides the gluon bath on which we will study flavor entanglement by imposing holographic prescriptions in open string geometries.

Another interesting information theoretic measure that  came into light in the study of two sided eternal $AdS$ black holes \cite{Maldacena:2001kr}, which is the bulk dual of well-known thermo-field double (TFD) states, is something called the holographic complexity. Although the study dates back prior to the discovery of holographic entanglement entropy, the particular suggestion of holographic complexity is more recent due to Susskind \textit{et al.} \cite{Susskind:2014rva, Stanford:2014jda, Brown:2015bva, Brown:2015lvg}.  The idea began to explain the growth of the size of Einstein-Rosen Bridges connecting two sides of the eternal black holes at time scales larger than what can be probed through the behaviour of entanglement entropy. Susskind \textit{et al.} suggested two different bulk calculations that can probe such a growth, which are famously known as \textit{Complexity equals Volume }(of the maximal volume slice connecting two sides of the black hole) and \textit{Complexity equals Action} (of the causal patch of the maximal volume slice, known as the Wheeler-DeWitt patch) proposals, defined as 
\begin{equation}
C_{V} = \left(\frac{V(\gamma)}{R G_{N}}\right), \qquad \qquad C_{A}=\frac{I_{WDW}}{\pi \hbar}~,
\end{equation}
where $R$ is the AdS radius, $V(\gamma)$ is the maximum volume slice connecting the two
boundaries of the black hole and $I_{WDW}$ is the action of the causal patch of the slice.

A  crucial point of these suggestions is that, they were proposed to be the holographic measure of a quantum information theoretic quantity known as computational complexity. Historically, the notion of complexity in computer science is the number of operations needed to implement a computational task. Now, evaluation of such a quantity primarily in quantum mechanics and eventually in quantum field theories, is itself a job that made people lean to the basics again and as it turns out, it all boils down to a problem of coupled quantum harmonic oscillators \cite{Chapman:2017rqy,Jefferson:2017sdb,Khan:2018rzm,Hackl:2018ptj}. The program tries to compute the cost of constructing a particular unitary operator made of a pre-decided set of reasonable quantum gates optimally\cite{Nielsen1, Nielsen2}. This unitary operator takes certain reference quantum state to the target state (with some tolerance involved) and the optimal cost function quantifying the optimal number of quantum gates needed to build this unitary gives the value of the corresponding circuit complexity.  There are different variants of these calculations \cite{Caputa:2017urj,Caputa:2017yrh, Abt:2017pmf} and we won't delve too deep inside that sea as our work doesn't really depend on them.  But it is certainly worth mentioning that baby steps have also been taken towards defining complexity in CFTs using the circuit complexity approach by considering the conformal transformation of the stress energy tensor generating the Virasoro group \cite{Caputa:2018kdj,Flory:2020eot,Erdmenger:2020sup,Caputa:2020mgb}.

Motivated by Susskind \textit{et al.}, another definition of holographic complexity has been proposed by Alishahiha \cite{Alishahiha:2015rta}, as the
 volume of the co-dimension one time-slice of the bulk geometry enclosed by the extremal codimension two Ryu-Takayanagi (RT) hypersurface
used for the computation of holographic EE. This is usually referred to as the subregion complexity 
\cite{Ben-Ami:2016qex, Carmi:2016wjl, Roy:2017kha, Bhattacharya:2018oeq} and the relation
between these two notions has been clarified in some recent works given in 
\cite{Agon:2018zso,Alishahiha:2018lfv,Takayanagi:2018pml}. This subregion complexity, which we calculate 
in this paper, is defined in a very similar way as, 
\be 
C_{V} = \frac{V_{RT}(\gamma)}{8\pi R G_N}~, \label{hsc}
\ee
where $V_{RT}$ denotes the
volume enclosed by the RT surface.    %However, a clear field theoretic description of holographic complexity is not yet known (see, however, \cite{Caputa:2017urj,
%Caputa:2017yrh}).
%Using the geometric approach of Nielsen to the quantum circuit model \cite{Nielsen1, Nielsen2}, the complexity of only some free field theory states have 
%been found to resemble the holographic complexity in \cite{Chapman:2017rqy,Jefferson:2017sdb,Khan:2018rzm,Hackl:2018ptj}. But, for interacting field theory this 
%is far from clear. Many different descriptions and proposals have been 
%given in the literature to relate holographic complexity to the fidelity susceptibility or the quantum Fisher information metric, the Bures metric 
%and so on \cite{MIyaji:2015mia, Alishahiha:2017cuk, Banerjee:2017qti}.
Similar in spirit to the case of HEE, we also look to compute holographic subregion complexity in the open string geometries applying (\ref{hsc}).

 This paper is organized as follows: In section \ref{sec2} we review the emergence of the kinematic open string  geometries and various energy conditions related to it. In section \ref{sec3} we study HEE  and its high-temperature limit for strip subsystems in the boundary of these asymptotically $AdS$ geometries in three, four and five dimensions. We also study HSC for the same setup. In section \ref{sec4} we carry out similar exercise for spherical subsystems. In section \ref{sec5} we explore the status of first law of entanglement thermodynamics in the probe flavor sector. Finally in section \ref{sec6} we summarize our results and  conclude with some discussion and open questions. We have also added a couple of Appendices for technical clarifications.

\section{Review of the open string metrics} \label{sec2}
 To set the stage, let us first review the emergence of open string geometries in our context. In the prototype version of the gauge/gravity duality \cite{Maldacena:1997re}, Maldacena considered low-energy stringy excitations in type IIB string theory in presence of a stack of almost coincident D3-branes. Since both the endpoints of the open strings in this setup must lie on the stack, the gauge theory  resulting from the massless excitations of the open strings on the D-branes consists of fields only in the adjoint sector. However, it is rather desirable that a highly successful framework like AdS/CFT  should give insights into QCD-like theories that describe our nature, comprising of  fundamental degrees of freedom as well along with gluonic sector. In the gauge/gravity duality framework, flavors are introduced in the gauge theory side by inserting additional D$p$-branes \cite{Karch:2002sh,DeWolfe:2001pq} in the supergravity background provided by the D3 stack or some other brane configuration.  For suitable values of $p$, the gauge theory living in the worldvolume of these \textit{flavor branes} decouples\footnote{The 't Hooft coupling constant  of the gauge theory vanishes.} and the strings stretching between the D3 stack and the flavor branes give rise to the desired flavor sector in the original gauge theory. Typically, introduction of $N_f$ flavor branes in a background sourced by $N_c$ number of (D or M) branes leads to $N_c^2$ gluon degrees of freedom coupled to $N_fN_c$ flavor degrees of freedom.

However, the insertion of these additional branes bears a couple of unwanted features as well. In the gravity side, the backreaction of these branes typically destroys the AdS asymptotics of the background.  Recast in the dual gauge theory language, the presence of the flavors renders the beta functions of the otherwise conformal gauge theory running \cite{Nunez:2010sf}. These features can however be circumvented in the following way: consider the supergravity partition function in presence of  $N_f$ number of D$p$-branes, schematically given by
 \be
\mathcal{Z}_{sugra+DBI} = \int D[\phi]D[g]D[\theta_i]D[F] ~ e^{-N_c^2 \mathcal{S}_{sugra}[\phi,g]}~ e^{-N_f N_c \mathcal{S}_{DBI}[g,\phi; \theta_i,F]}~,
\ee
 where $\{\phi,G\}$ are supergravity fields, $\{\theta_i,F\}$ are the fields of the worldvolume theory. In the limit $N_c^2 \rightarrow \infty$, $N_c N_f \rightarrow \infty$, we can perform saddle point approximation for both the theories leading to the classical partition function
  \be
\mathcal{Z}_{sugra+DBI}^{(classical)} =  e^{-N_c^2 \mathcal{S}_{sugra}^{(0)}- N_f N_c \mathcal{S}_{DBI}^{(0)} - N_f^2 \mathcal{S}_{back-reac}^{(1)} + \mathcal{O}(N_f/N_c)}~.
\ee
Here $\mathcal{S}_{sugra}^{0}$ and $\mathcal{S}_{DBI}^{0}$ are the on-shell values of the actions corresponding to the classical minima of their respective theories. The term $N_f^2 \mathcal{S}_{back-reac}^{(1)}$ captures the backreaction of the brane on the supergravity background. So clearly in the limit $N_c^2 \rightarrow \infty$, $N_c N_f \rightarrow \infty$, $N_f/N_c \ll 1$, the backreaction drops out. This is the so-called \textit{probe} limit where the branes do not backreact on the background geometry. In the dual gauge theory,  this amounts to the so-called  \textit{quenched} approximation where the quarks are classical objects and their dynamics do not affect the strongly interacting background provided by the gluons. Now using holographic principle, this dynamics of the quarks can be studied by considering the dynamics of the probe flavor branes in the supergravity background, given by the Dirac-Born-Infeld(DBI) action\footnote{The action we consider here (\ref{DBI}) corresponds to a single D$p$ brane, for which the worldvolume gauge field is Abelian. For a stack of $N_f$ coincident branes, the gauge field is $U(N_f)$ valued and the field strength in no longer gauge invariant. However, in the discussions to follow, we will  primarily be studying the Abelian DBI theory without going into the technicalities of non-Abelian DBI theory \cite{Tseytlin:1997csa}. That is to say, either we will consider a single probe brane or a stack of branes with finite spacing among themselves with the additional  assumption that the gauge fields living on these branes are exactly identical. This amounts to adding a $U(1)^{N_f}$ symmetric flavor sector as opposed to a $U(N_f)$ sector.}   

\begin{equation}
S_{\text{D}p} = \tau_{p} \int_{\mathcal{M}_{p+1}} d^{p+1}y~ e^{-\Phi}~ \sqrt{-det\left[P\left(G_{ab} + B_{ab}\right) + 2\pi \alpha' F_{ab}\right]} ~, \label{DBI}
\end{equation}
 where $\tau_p = (2\pi)^{-p}g_s^{-1} \alpha'^{-(p+1)/2}$ is  the brane tension,  $P[G_{ab} + B_{ab}]$ is the pull-back of the background metric and the B-field onto the worldvolume $\mathcal{M}_{p+1}$, $F_{ab}$ is the field strength corresponding to the gauge field living in the worldvolume. The dynamical fields of the DBI theory are thus the transverse scalars $\{\theta_i\}$ and the worldvolume gauge field $A_{a}$ resulting from the transverse and longitudinal oscillations of the open strings on the branes respectively. On studying fluctuations around the classical saddle of the DBI-theory,
 
 \begin{equation}
 \theta_{i} = \theta_i^{(0)} + \varphi_i ~, ~~~~ A_{a} = A_{a}^{(0)} + \mathcal{A}_{a}~,
 \end{equation}
 it turns out that the kinetic terms for the fluctuations take the following form
\begin{eqnarray}
S_{scalar} & = & - \frac{\kappa}{2} \int d^ay\left(\frac{det G}{det \mathcal{S}} \right)^{1/4} \sqrt{-det \mathcal{S}} \,  \mathcal{S}^{ab} \, \partial_a \varphi^i \,  \partial_b \varphi^i    , \label{fscalar} \\
	S_{ vector} & = & - \frac{\kappa}{4}  \int d^a y \left(\frac{det G}{ det \mathcal{S}} \right)^{1/4} \sqrt{-det \mathcal{S}}  \mathcal{S}^{ab}\mathcal{S}^{cd} \, \mathcal{F}_{ac} \mathcal{F}_{bd} . \label{fvector} 
	\end{eqnarray} 
 where
 \begin{equation}\label{osm}
 \mathcal{S}_{ab}= P[G]_{ab} - \left(F^{(0)}~. P[G]^{-1}~.F^{(0)}\right)_{ab}~.
 \end{equation}
 is the open string metric(osm henceforth). We have shown only the kinetic parts of the fluctuation Lagrangian; since other potential terms will not affect our discussion for now. The Lagrangian density corresponding to (\ref{fscalar}) and (\ref{fvector}) can be written in a more canonical form:
	\begin{eqnarray}
	&& \sqrt{- det \tilde{\mathcal{S}} }\ \tilde{\mathcal{S}}^{ab} \left(\partial_a \varphi \right)  \left(\partial_b \varphi \right) \ , \quad {\rm and} \quad  \sqrt{- det \tilde{\mathcal{S}}} ~ \tilde{\mathcal{S}}^{ab} \tilde{\mathcal{S}}^{cd} \F_{ac} \F_{bd} \ , \label{canokin} \\
	&& {\rm where} \quad \tilde{\mathcal{S}}_{\rm ab} = \Omega  \mathcal{S}_{\rm ab}  \label{confosm}
	\end{eqnarray}
	and $\Omega$ needs to be determined for each dimensions, separately. Since conformal rescaling does not change the causal structure of the spacetime, it is a matter of choice to pick between $\mathcal{S}_{ab}$ and  $\mathcal{\tilde{S}}_{ab}$ for the following discussion. We will simply work with (\ref{osm}).
	
The kinetic terms in  (\ref{fscalar}) and (\ref{fvector}) suggest that the fluctuations of the brane do not perceive the pullback metric, rather they see the osm.\footnote{Note that had we turned on a Maxwell field $F_{ab}$ in a spacetime $g_{ab}$, the scalar and vector fluctuations in this background will always perceive $g_{ab}$. So osm is completely inherent to brane fluctuations only.} Clearly the osm differs from the pullback geometry in presence of a non-vanishing field strength. Also the osm does not follow from extremization of some action, rather it emerges kinematically from the background metric and certain field configuration in the worldvolume. One remarkable feature of this metric is that, even if the background metric does not have a horizon, due to the second term in (\ref{osm}), one can engineer a horizon in the osm by suitably choosing the field strength configuration in the worldvolume. In particular, this is obtained by exciting an electric field in the worldvolume, which sets the horizon of the geometry and hence an effective temperature $T_{eff}$ for the  brane fluctuations.  
 
Thus the introduction of $N_f$ number of D$p$-branes in the limit $N_c^2 \rightarrow \infty$, $N_c N_f \rightarrow \infty$, $N_f/N_c \ll 1$  leads to conjecturing a new duality between gravity in  open string geometries and the physics of the flavors in the dual gauge theory. This duality has been exploited to study thermodynamics of the flavor fields \cite{Banerjee:2015cvy}, chaos in the flavor sector \cite{Banerjee:2018kwy} and in numerous other contexts. Also a comparative study of the causal structures between these open string geometries and the standard black hole solutions in Einstein gravity has been carried out extensively in \cite{Banerjee:2016qeu}. Being motivated by the similarities between the two, the authors in \cite{Banerjee:2016qeu} looked for matter field configurations which may yield these geometries when coupled to Einstein gravity. However, it turned out that the resulting stress tensor becomes pathological, in the sense that it violates one of the energy conditions in GR, depending on the dimensions. The main focus of this article is to study entanglement entropy and complexity in the flavor sector holographically, applying the standard RT prescriptions to open string geometries. These exercises are thus  expected to capture the robustness of RT proposal to violation of energy conditions and to explore their compatibility with non-Einstein solutions.

\subsection{Open string geometry in various dimensions}
%%%%%%%%%%%%%%%%%%%%%%%%%%%
In this section, we will give explicit instances of open string geometries in various dimensions, starting from pure-$AdS$ background. We start with the $AdS_{2+1}$ metric in the following form:
\begin{equation}
ds^2= \frac{1}{z^2} \left[ - dt^2 + dx^2 + dz^2 \right] ~,  \label{ads3def}
\end{equation}
where the $AdS$ radius has been set to unity for the rest of discussion. The conformal boundary is located at $z \to 0$ and the infrared of the geometry is located at $z \to \infty$. The gravity fluctuations in this background does not perceive any temperature. In the dual field theory, this corresponds to a gluon bath at zero temperature.

Now we will consider introducing ``space-filling" probe branes in the $AdS_3$-background. We will make the assumption that such space-filling embedding exists, without going into the details of the brane configurations . This, for the most part of our purpose, is a simplifying assumption that does not necessarily cost any physical information. Now to introduce an event horizon in the osm, we will excite the following gauge potential: 
\begin{eqnarray}
A_x = - E t + a_x(z) \quad {\rm with} \quad  F = dA \ . \label{gauge}
\end{eqnarray}
The physics of this fundamental matter sector is quite intuitive: Since we applied an electric field\footnote{This electric field couples to the flavor sector only.}, there will be pair-creation even in the absence of explicit charge density and this will drive a flavor-current. The corresponding current, denoted by $j \sim \left( \partial \mathcal{L}_{\rm DBI} / \partial a_x'\right)$, is essentially given by the first integral of motion for the field $a_x(z)$. See {\it e.g.}~\cite{Karch:2007pd, Albash:2007bq,Erdmenger:2007bn} for more details on a representative example of embedding D$7$-brane in AdS$_5\times S^5$-background.

Now, using the definition in (\ref{osm}), for the background in (\ref{ads3def}) and the gauge field in (\ref{gauge}) the corresponding osm in three dimensions is calculated to be:
\begin{eqnarray}
ds_{\rm osm}^2 & = & -\frac{1}{z^2}\left(1-\frac{z^4}{z_h^4}\right)d\tau^2+\left(\frac{1}{z^2}+\frac{1}{z_h^2}\right) dx^2 + \frac{1}{z^2}\left(\frac{1}{1-\frac{z^2}{z_h^2}}\right)dz^2 , \label{osm3} \\
d\tau & = &  dt - \frac{E j z^3}{\sqrt{\left( 1- E^2 z^4  \right) \left( 1-j^2 z^2  \right)}} dz \ , \label{tautdef}
\end{eqnarray}
with
\begin{equation}
E = \frac{1}{z_h^2} \ , \qquad  j = \frac{1}{z_h} = \sqrt{E} \ .
\end{equation}
With reference to (\ref{confosm}), also note that
\begin{eqnarray}
\tilde{\mathcal{S} }= \Omega \mathcal{S} \ , \quad \Omega = \left( 1 + \frac{z^2}{z_h^2} \right)^{-1} \ . \label{confosm3}
\end{eqnarray}
Clearly, the osm in (\ref{osm3}) inherits a structure similar to a black hole geometry, with an effective Hawking temperature:
\begin{equation}
T_{\rm eff} =  \frac{E^{1/2}}{\sqrt{2}\pi} \ . \label{osmtemp}
\end{equation}
This is the temperature the brane fluctuations perceive. So in the putative dual field theory, the flavor sector is now at a finite temperature $T_{eff}$ while the gluon sector is at zero temperature. However, in the probe limit, the heat flow from the flavor to the gluon sector is $\mathcal{O}(N_f/N_c)$ suppressed and hence both the  sectors are in thermodynamic equilibrium of their own. \footnote{Strictly speaking, the flavor sector is in a non-equilibrium steady state(NESS), owing to the current flow induced by the electric field.}\vspace{4mm} \hfill \break 
Similarly, the open string metrics in higher dimensions are obtained to be
\begin{align}
ds_{(4)}^2 &=  \frac{1}{z^2}\left[-(1-E^2z^4)dt^2 + \frac{dz^2}{1-E^2z^4} + \left(dx_1^2 + dx_2^2\right)\right] , \label{osm4} \\
ds_{(5)}^2 &=   \frac{1}{z^2}\left[-(1-E^2z^4)dt^2 + \frac{dz^2}{1-E^3z^6} + \frac{1-E^2z^4}{1-E^3z^6}dx_1^2 +  \left(dx_2^2 + dx_3^2\right) \right]~.  \label{5osm}
\end{align}
Note that in four dimensions, there is an accidental isotropy between the directions longitudinal and transversal to the applied electric field, which is however absent in five dimensions. Qualitatively, this anisotropy is similar to the one observed in \cite{Narayan:2012ks} resulting from an energy current, whereas in our case it emerges due to the charge current driven by the electric field.  This anisotropy will eventually play a crucial role in the choice of strip-like subregion in the boundary for RT analysis \cite{Narayan:2012ks,Narayan:2013qga,Mukherjee:2014gia,Mishra:2016yor}.
%%%%%%%%%%%%%%%%%%%%%%%%%
\subsection{Energy Conditions}
%%%%%%%%%%%%%%%%%%%%%%%
Even though open string metrics emerge kinematically from brane configurations, they bear stark resemblance to  black holes in Einstein gravity over various aspects \cite{Banerjee:2016qeu}.  It is therefore worth exploring whether there exists \textit{sensible} matter fields which, when  coupled to Einstein gravity may give rise to these geometries. To start with, let us choose the notation: we use $ G_{\mu\nu} $ to denote the corresponding Einstein-tensor evaluated from the given open string metric $\mathcal{S}_{\mu\nu}$. The {\it equation} we pretend solving is the following:
 \begin{eqnarray}
 G_{\mu\nu} + \Lambda \mathcal{S}_{\mu\nu} = \T_{\mu\nu} \ , \label{einpretend}
 \end{eqnarray}
 where $\Lambda = - d(d-1) / 2$ is the cosmological constant in asymptotically $AdS_{d+1}$ background, and $\T_{\mu\nu}$ is the stress tensor of the putative matter field. With this $\T_{\mu\nu}$, we will investigate the following energy conditions:\\
 
 \noindent {\bf (i) Null Energy Condition (NEC):} This implies that for every future pointing null vector, the matter density observed by the corresponding observer is non-negative. For a given $\T_{\mu\nu}$ and any null vector $n^\mu$, the null energy condition imposes: $\T_{\mu\nu} n^\mu n^\nu \ge 0$. For the discussions to follow, we will choose a generic null vector of the form $n^{\mu}= \left\{ n_1(z),n_2(z),0,\ldots\right\}$, such that $g_{\mu\nu} n^{\mu}n^{\nu}=0$.\\
 
 \noindent {\bf (ii) Weak Energy Condition (WEC):} It implies that for every future pointing timelike vector, the matter density observed by the corresponding observer is non-negative. For a given $\T_{\mu\nu}$ and any timelike vector $t^\mu$, the weak energy condition imposes: $\T_{\mu\nu} t^\mu t^\nu \ge 0$. Again we will choose generic timelike vector of the form $t^{\mu}= \left\{t_1(z),t_2(z),0,\ldots\right\}$, with $g_{\mu\nu} t^{\mu}t^{\nu}=-1$.\\
 
With these ingredients, let us now explore the energy conditions in various dimensions.
In three dimensions
\bea
\T_{\mu\nu} n^\mu n^\nu &=& - \frac{2 E~ n_1(z)^2}{(1+ E z^2)^2}~,\\
\T_{\mu\nu} t^\mu t^\nu &=& \frac{2 E\left(E z^4 -t_1(z)^2\right)}{(1+ E z^2)^2}~.
\eea
In four dimensions
\bea
\T_{\mu\nu} n^\mu n^\nu &=& 0~,\\
\T_{\mu\nu} t^\mu t^\nu &=& - E^2 z^4~.
\eea
Finally in five dimensions
\bea
\T_{\mu\nu} n^\mu n^\nu &=&\frac{2 E^2 z^2\left(6+ E z^2\left(2+E z^2\right)\right)^2}{\left(1+ 2 E z^2 + 2 E^2 z^4 + E^3 z^6\right)^2} ~n_1(z)^2~,\\
\T_{\mu\nu} t^\mu t^\nu &=& -\frac{E^3 z^6\left(9+E z^2\left(23 +Ez^2 \left(18+ Ez^2\left(3+ Ez^2\right)\right)\right)\right)}{\left(1+ E z^2\right)^2\left(1+  E z^2 +  E^2 z^4 \right)}~.
\eea
To summarize, the four and five dimensional opens string metrics always violate WEC, but preserve NEC. On the other hand, in three dimensions NEC is always violated, while the violation of WEC is subtle. These violations imply that there is no area-increase theorem for the osm horizon area, and consequently, we cannot identify this area with thermal entropy in the putative dual field theory \cite{Banerjee:2016qeu}. In this article, we will try to capture the imprints of these violations on HEE and HSC computed following the standard holographic prescriptions.
%%%%%%%%%%%%%%%%%%%%%%%%%%%%%%%%%
\section{HEE and HSC for strip subregion}  \label{sec3}
\subsection{Holographic Entanglement Entropy}
As mentioned in section \ref{sec1}, Ryu-Takayanagi conjecture provides us with a way of measuring entanglement between parts of the boundary dual to the bulk using bulk minimal-surface prescription. There are numerous extensions of this proposal \textit{e.g.} Hubeny-Rangamani-Takayanagi (HRT) prescription for time dependent(non-static) case \cite{Hubeny:2007xt}, generalized gravitational entropy \cite{Lewkowycz:2013nqa, Faulkner:2013ana}, quantum maximin and quantum extremal surface prescriptions \cite{Engelhardt_2015} that deal with various kind of corrections over the RT prescription. In this paper, we nevertheless restrict ourselves to the original RT prescription for static spacetimes. For application of Ryu-Takayanagi formula \eqref{HEE} in open string geometries, we will first look at subsystems with shape of straight strips, having one of their spatial extension $l$ narrower than the rest. Although the longer directions are in principle unbounded, in practice we will always put a regulator $L(\gg l)$ to avoid a divergence. We will treat the metrics (\ref{osm3}),(\ref{osm4}) and (\ref{5osm}) as perturbations over pure $AdS$ spacetimes wherein the electric field will play the role of the perturbation parameter.
\subsubsection{$(2+1)$ dimensions :}
The first example deals with $(2+1)$ dimensional osm given by,
	\begin{equation}
		ds_{(3)}^2 = \frac{1}{{z}^2}\left[-(1-E^2{z}^4) dt^2 + (1+E {z}^2) d{x}^2 + \frac{d{z}^2}{1-E{z}^2} \right]~.
	\end{equation}
	In this coordinates, ${x} \in [-\frac{l}{2}, \frac{l}{2}]$. The RT surface in this 3-dimensional scenario is essentially the geodesic through the bulk anchored between the endpoints of the strip. As usual, we will choose a constant time slice and let ${x} = {x}(z)$. Then the length of this geodesic can be written as:
	\begin{equation}
		\mathcal{A} = 2\int_{\epsilon}^{{z}_*} \frac{d{z}}{{z}}\sqrt{\frac{1}{1-E{z}^2} + (1+E{z}^2) {x}^{\prime 2}({z})}~. \label{area3}
	\end{equation}
	Here, $\epsilon$ is a cutoff introduced to protect the integral from an UV divergence near ${z}\to 0$ and ${z}_*$ denotes the turning point of the curve. To apply the RT formula, the integral above needs to be extremized, leading to the Euler-Lagrange equation

	\begin{align}
	\nonumber	&\frac{(1+E z^2)x^{\prime}(z)}{z\sqrt{\frac{1}{1-E z^2} + (1+E z^2)x^{\prime}(z)^2}} = b ~,\\ 
		\text{or, }~~ &x^{\prime}(z)^2 = \frac{b^2z^2}{(1-E z^2)\left[(1+E z^2)^2 - (1+E z^2)b^2z^2\right]}~.
    \end{align}
	The constant $b$ can be determined from the condition that at the turning point $z = z_*$, $\frac{dz}{dx} = 0$, therefore, 
	\be
    b^2 = \frac{(1+E z_*^2)}{z_*^2} ~,
    \ee
	and consequently,
	\begin{align*}
		\int_{0}^{\frac{l}{2}}dx &= \int_{0}^{z_*} \frac{\frac{z}{z_*}}{\sqrt{\frac{1+E z^2}{1+E z_{*}^2}-\frac{z^2}{z_*^2}}\sqrt{1-E^2z^4}}~dz~,\\
		\text{or, }\qquad\frac{l}{2} &= z_*\int_{0}^{1}\frac{y \sqrt{1+\lambda z_{*}^2} ~dy}{\sqrt{1-y^2}\sqrt{1-E^2z_*^4 y^4}}~,
	\end{align*}
	where $y = \frac{z}{z_*}$. Now we will work in a regime where $E$ \footnote{Note that we are working in units where $E$ is dimensionless.} is very small and perform a power series expansion in $E$, keeping terms upto $\mathcal{O}(E^2)$, leading to an approximate relationship between the turning point and the width of the strip:
	\begin{equation}\label{turnpt3}
		z_{*} = \frac{l}{2}\left[1 - \frac{1}{8}E l^{2} +\frac{73}{1920}E^{2} l^{4}\right] + \mathcal{O}\left(E^{3}\right)~.
	\end{equation}
	Note that for $E = 0$: $z_* = \frac{l}{2}$, which is a familiar relationship for pure $AdS_3$.
	 The extremal path length in $\eqref{area3}$ can now be evaluated:
	\begin{align*}
		\mathcal{A} = 2\int_{\epsilon/z_*}^{1}\frac{dy}{y}\sqrt{\frac{1+ E z_*^2 y^2}{(1-E z_*^2 y^2)(1 - y^2)}}~.
	\end{align*}
	Performing a series expansion in powers of $E$, one readily finds:
	\begin{equation}\label{finalarea3}
		\mathcal{A} = 2\left[\log\left(\frac{l}{\epsilon}\right) + \frac{1}{8}E l^2 - \frac{11}{960}E^2 l^4 \right]~.
	\end{equation}
	The first term in this expression is the ground state result and is in agreement with  results known from calculation in pure $AdS_3$, whereas subsequent terms denote corrections coming at various orders due to excitation. \vspace{2mm} \newline
The change in entanglement entropy over pure $AdS_3$ state is thus
	\begin{equation}
	    \Delta S_{(3)} = \frac{E~l^2}{4G_{eff}^{(3)}}\left[\frac{1}{4} - \frac{11}{480}El^2 \right] + \mathcal{O}(E^3)~. \label{hee3}
	\end{equation}
It is instructive to compare the result \eqref{hee3} with that for a BTZ black hole. For a non-rotating BTZ black hole the HEE is given by \cite{Hubeny:2007xt}

\begin{equation}\label{SBTZ}
    S_{BTZ} = \frac{1}{2G_N^{(3)}} \log\left[ \frac{2}{\epsilon \sqrt{m}}\sinh \left({\frac{\sqrt{m} l}{2}} \right) \right]~,
\end{equation}
where $m$ sets the BTZ temperature  $T_{BTZ} = \frac{\sqrt{m}}{2\pi}$. Performing a series expansion in $m$ and ignoring ground state contribution, one finds
\begin{equation}
    \Delta S_{BTZ} = \frac{m l^2}{4G_N^{(3)}}\left[ \frac{1}{12} - \frac{ml^2}{1440} \right] + \mathcal{O}\left(m^3\right)~.
\end{equation}
An order by order comparison\footnote{In order to compare, we set the two black holes at equal temperature. This relates $E=m/2$.} with \eqref{hee3} gives
\begin{eqnarray}
\nonumber \Delta S^{(1)}_{(3)} &= \frac{3}{2} \Delta S_{BTZ}^{(1)}~,\\
\Delta S^{(2)}_{(3)} &= \frac{33}{4} \Delta S_{BTZ}^{(2)}~.\label{osmbtzst}
\end{eqnarray}
\subsubsection{$(3+1)$ dimensions :}
The 4-dimensional open string metric is given by:
	\begin{equation}\label{Ads4}
	    ds_{(4)}^2 = \frac{1}{z^2}\left[-(1-E^2z^4)dt^2 + \frac{dz^2}{1-E^2z^4} + \left(dx_1^2 + dx_2^2\right)\right]~.
	\end{equation}
	As for the boundary strip region, we will choose $-\frac{l}{2} \leq x_1 \leq \frac{l}{2}$, $0 \leq x_2 \leq L$ and parametrize the co-dimension 2 surface as $x_1 = x(z)$,\footnote{The isotropy along the boundary spatial directions allows us to pick any of the two directions.} leading to the following area integral:
	\begin{equation}\label{area4}
	    \mathcal{A} = L\int_{\epsilon}^{z_*}\frac{dz}{z^2}\sqrt{x'(z)^2 + \frac{1}{1 - E^2z^4}}~.
	\end{equation}
	The extremization of this integral will as before lead to a relationship between the width $l$ of the strip and the turning point $z_*$, which in this case takes the form
	\begin{equation}
	    \frac{l}{2} = z_* \int_0^1 \frac{y^2dy}{\sqrt{(1-y^4)(1-E^2z_*^4y^4)}}~.
	\end{equation}
	The integral has an analytic solution in the regime $Ez_*^2 < 1$, which is also the regime of our interest, yielding
	\begin{equation}
	    \frac{l}{2} = \Hypergeometric{2}{1}{\frac{1}{2}, \frac{3}{4}}{\frac{5}{4}}{E^2z_*^4}b_0 z_*~.
	\end{equation}
	From this, the turning point can be approximately written as:
	\begin{equation}
	  		z_* = \bar{z_*}\left[1 - \frac{3}{10}E^2 \bar{z_*}^4 +\frac{11}{40}E^4 \bar{z_*}^8\right] + \mathcal{O}\left(E^6\right)~,
	\end{equation}
	where, $\bar{z_*} = \frac{l}{2b_0}$ with $b_0 = \frac{\sqrt{\pi}\Gamma(\frac{3}{4})}{\Gamma(\frac{1}{4})}$. We can now evaluate the area integral \eqref{area4} perturbatively in $E\bar{z_*}^2$, leading to 
	\begin{equation}
	    \mathcal{A} = \mathcal{A}_0 + 2 L\left[\frac{1}{40}\frac{E^2l^3}{b_0^2} - \frac{1}{3200}\frac{E^4l^7}{b_0^6} \right]+ \mathcal{O}\left(E^6\right)~,
	\end{equation}
	where we denote the ground state $(AdS_4)$ area as $\mathcal{A}_0$ and the change in HEE is then
	\begin{equation}{\label{ads4}}
	    \Delta S_{(4)} = \frac{E^2 L l}{4G_{eff}^{(4)}}\left[\frac{1}{20}\frac{l^2}{b_0^2} - \frac{1}{1600}\frac{E^2l^6}{b_0^6} \right] + \mathcal{O}\left(E^6\right)~.
	\end{equation}
	
\subsubsection{$(4+1)$ dimensions :}
The open string metric in $5$-dimensions is:
	\begin{equation}
	    ds^2_{(5)} = \frac{1}{z^2}\left[-(1-E^2z^4)dt^2 + \frac{dz^2}{1-E^3z^6} + \frac{1-E^2z^4}{1-E^3z^6}dx_1^2 + \left(dx_2^2 + dx_3^2\right) \right]~.
	\end{equation}
	As before, the electric field has been chosen to be applied along $x_1$. However, a crucial difference in five dimensions as compared to four dimensions is the breaking of isotropy between the directions longitudinal and transversal to the electric field. This leads to two physically distinct choice for the strip --- (a) \textbf{Strip along the electric field}. In this case the strip width is chosen perpendicular to the electric field direction $x_1$, so the flavor current flows along the strip. (a) \textbf{Strip orthogonal to the electric field}. In this case the strip width is chosen along $x_1$ and hence the strip extends perpendicular to the flow. Hence in this case the entire current flows across the strip and there is constant current exchange with the adjacent region.
	
	\textbf{(a) Strip along the electric field:}
	Let us  consider a strip like subsystem such that $-\frac{l}{2} \leq x_2 \leq \frac{l}{2}$ and $0 \leq x_1, x_3 \leq L$. With the parametrization $x_2 = x(z)$  the area integral becomes:
	\begin{equation}\label{area5perp}
	    \mathcal{A} = 2 L^2\int_{\epsilon}^{z_*}\frac{dz}{z^3}\sqrt{x'(z)^2 + \frac{1}{1-E^3z^6}}\sqrt{\frac{1-E^2 z^4}{1- E^3z^6}}~.
	\end{equation}
	To solve for the turning point $(z_*)$ one needs to solve: 
	\begin{equation*}
	    \frac{l}{2} = z_*\int_0^1 \frac{y^3 dy}{\sqrt{\frac{\omega^2(z)}{\omega^2(z_*)} - y^6}\sqrt{1-E^3z_*^6y^6}}~,
	\end{equation*}
	where we've denoted $\omega(z) = \sqrt{\frac{1-E^2 z^4}{1- E^3z^6}}$. Following a perturbative expansion we obtain the following expression for $z_*$ approximated up to second non-trivial order
\begin{equation}
	    z_* = \frac{\bar{z}_*}{1 - \frac{1}{2}E^2\bar{z}_*^4 \frac{I_1}{b_0} + \frac{1}{2}E^3\bar{z}_*^6 \frac{I_2}{b_0}}~,
\end{equation}
where as before, $\bar{z}_* = \dfrac{l}{2b_0}$ and $b_0, I_1$ \& $I_2$ are beta integrals which are listed below. The area integral is similarly calculated to obtain:
	\begin{equation}
	    \mathcal{A} = \mathcal{A}_0 + 2L^2\left[ -\frac{1}{2}E^2\bar{z}_*^2\left(\frac{J_1}{2} + \frac{2a_0}{b_0}I_1\right) + \frac{1}{2}E^3\bar{z}_*^4 \left(3b_0 + \frac{2a_0}{b_0}I_2\right) \right]~.
	\end{equation}
	Let us list here the several beta integrals we encountered
	
	\begin{align}\label{betadef1}
	    &b_0 = \int_0^1 \frac{y^3 dy}{\sqrt{1-y^6}} = \frac{B(\frac{2}{3}, \frac{1}{2})}{6}~, \nonumber \\ 
	    &a_0 = \int_0^1 \frac{dy}{y^3\sqrt{1 - y^6}} = -\frac{b_0}{2}~,\nonumber \\
	    &I_1 = \int_0^1 \frac{y^3(1-y^4)dy}{(1-y^6)^{\frac{3}{2}}}  = \frac{1}{9}B\left(\frac{1}{3}, \frac{1}{2}\right) - \frac{b_0}{3} \nonumber ~,\\
	    &J_1 = \int_0^1 \frac{y(1+y^2-2y^6)dy}{(1-y^6)^{\frac{3}{2}}}   = \frac{B(\frac{1}{3}, \frac{1}{2})}{6} + I_1 \nonumber~, \\
	    &I_2 = \int_0^1 \frac{y^3(1+y^6)dy}{\sqrt{1-y^6}} = \frac{11}{7}b_0~.
	\end{align}
	After simplification, we finally get 
	\begin{equation}
	    \mathcal{A} = \mathcal{A}_0 + 2 L^2\left[-\frac{E^2l^2}{12}\frac{B(\frac{1}{3}, \frac{1}{2})}{4b_0^2} + \frac{5~E^3l^4}{7} \frac{B(\frac{2}{3}, \frac{1}{2})}{16b_0^4} \right]~.
	\end{equation}
	So the change in HEE, therefore, is: 
	\begin{equation}
	    \Delta S_{\left(5\parallel\right)} = \frac{E^2 L^2 l}{4G_{eff}^{(5)}}\left[-\frac{B(\frac{1}{3}, \frac{1}{2})}{6}\frac{l}{4b_0^2} + \frac{5}{7} B(\frac{2}{3},\frac{1}{2})\frac{E l^3}{8b_0^4}\right] + \mathcal{O}(E^4)~.
	\end{equation}
	Note that the successive orders are suppressed by factors proportional to $E$, so the leading order  was expected to occur at $\mathcal{O}(E)$. However, due to the metric structure this term is absent rendering the leading order contribution negative.
	
	\textbf{(b) Strip orthogonal to the electric field:}
	Next, we consider the strip-width along the direction of the electric field (i.e. $-\frac{l}{2} \leq x_1 \leq \frac{l}{2}$ and $0 \leq x_2, x_3 \leq L$). With the parametrization $x_1=x(z)$, the area integral  in this case becomes
	\begin{equation}\label{area5par}
	    \mathcal{A} = 2 L^2\int_{\epsilon}^{z_*}\frac{dz}{z^3}\sqrt{\frac{1- E^2z^4}{1-E^3 z^6}~x'(z)^2 + \frac{1}{1-E^3z^6}}~.
	\end{equation}
	Rest of the calculations are similar as before. The turning point $z_*$ is related to the strip-width $l$ as
	\begin{equation}
	    z_* = \frac{\bar{z}_*}{1 - \frac{1}{2}E^2\bar{z}_*^4\frac{\tilde{I}_1}{b_0} + \frac{1}{2}E^3\bar{z}_*^6}~,
	\end{equation}
	which is true up to $\mathcal{O}(E^3l^6)$ and the co-efficient $\tilde{I}_1$ is expressible in terms of Beta function. A perturbative analysis for the area integral \eqref{area5par} results in
	\begin{equation}
	    \mathcal{A} = \mathcal{A}_0 - L^2~E^2\bar{z}_*^2 \left(\tilde{J}_1 + \frac{2a_0}{b_0}\tilde{I}_1 \right) + \mathcal{O}(E^4)~,
	\end{equation}
	where as always, $\mathcal{A}_0$ denotes the ground state (pure $AdS_5$) result, $a_0$ and $b_0$ represent the same integrals defined in \eqref{betadef1}, and hence $2a_0 = - b_0$. So the new coefficients are related by
	\begin{equation*}
	    \tilde{J}_1 - \tilde{I}_1 = \int_0^1 \frac{y^7dy}{\sqrt{1-y^6}} = \frac{B(\frac{1}{3}, \frac{1}{2})}{15}~.
	\end{equation*}
	Therefore, the area of the RT surface becomes
	\begin{equation}
	    \mathcal{A} = \mathcal{A}_0 - L^2~\frac{E^2l^2}{15}\frac{B\big(\frac{1}{3}, \frac{1}{2}\big)}{4b_0^2}+\mathcal{O}(E^4)~,
	\end{equation}
	and hence, the holographic entanglement entropy suffers the change
	\begin{equation}
	    \Delta S_{\left(5\perp\right)} = - \frac{E^2 L^2 l}{4G_{eff}^{(5)}} \frac{B\big(\frac{1}{3}, \frac{1}{2}\big)}{15}\frac{l}{4b_0^2} + \mathcal{O}(E^4)~.
	\end{equation}
So the leading non-vanishing contribution to the change in HEE is negative again, as before. But there is a surprising cancellation at the subleading order resulting from the expansion of the turning point and the area integral. So in this, the subleading contribution occurs at $\mathcal{O}(E^4)$, whereas for the parallel case, it occurs at $\mathcal{O}(E^3)$.

\subsection{Numerical results:}
So far we have restricted ourselves to the narrow strip domain $\left(l \ll z_h\right)$,  as to remain within the perturbative regime where analytic computations are possible. In this section, we carry out some numerical exploration to check whether the perturbative results are trustworthy at all. This is further motivated by the results in five dimensions which a priori looks surprising.
\par To do the numerics we choose $L = 1$ and $z_h = \sqrt{10}$ i.e. $E = \frac{1}{100}$. We then choose a range of values for the turning point $z_*$. For the selected range we find out the corresponding strip-width $l$ and the area of the RT surface for $d = 2, 3, 4$. We calculate the area difference from the ground state $(E=0)$ and plot this difference against the $l$ values.
\par
\begin{figure}
	\begin{subfigure}{0.49\textwidth}
		\centering
		\includegraphics[width=\textwidth]{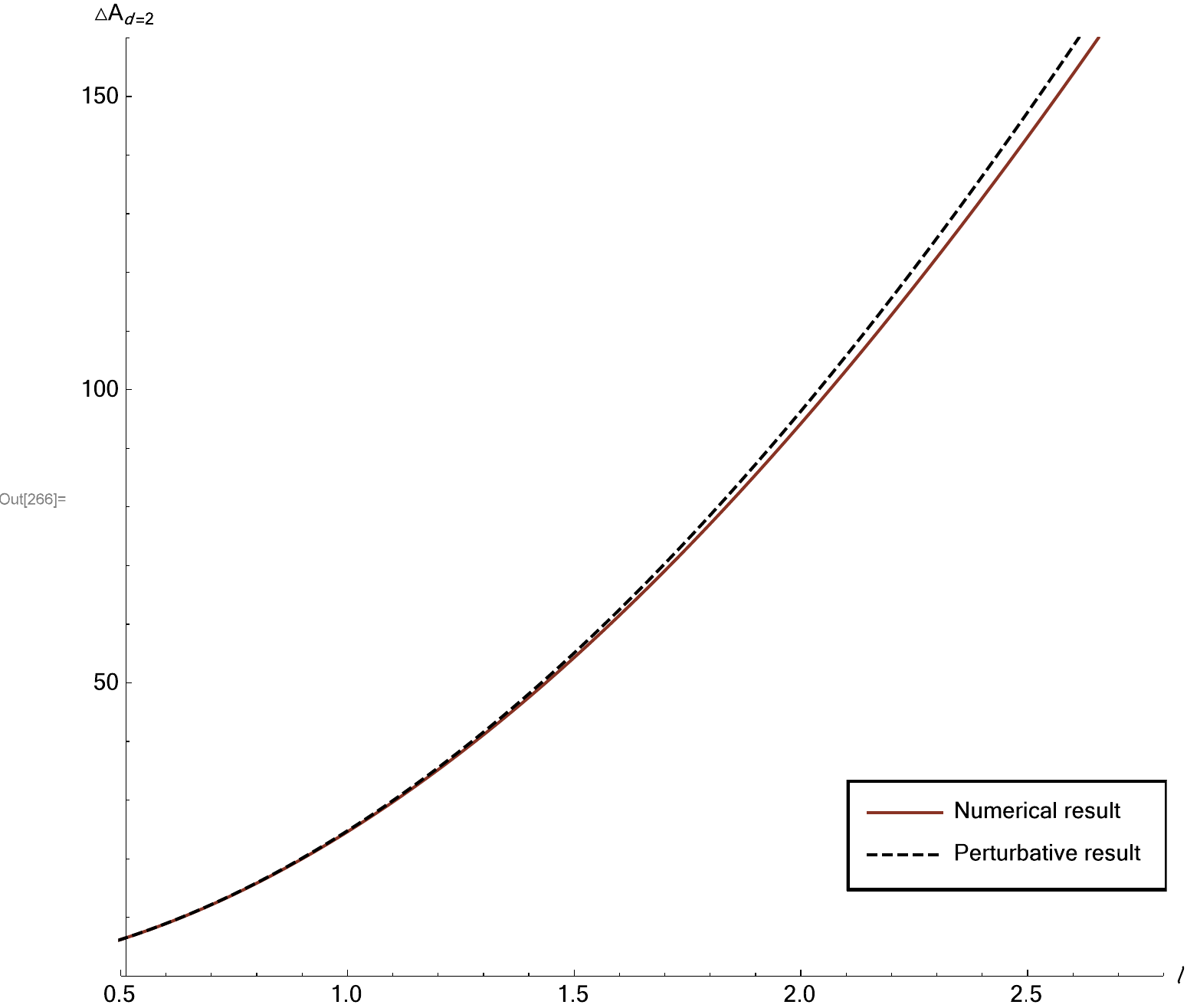}
		\caption{}
		\label{fig:NEE3D}
	\end{subfigure}
	\hfill
	\begin{subfigure}{0.49\textwidth}
		\centering
		\includegraphics[width=\textwidth]{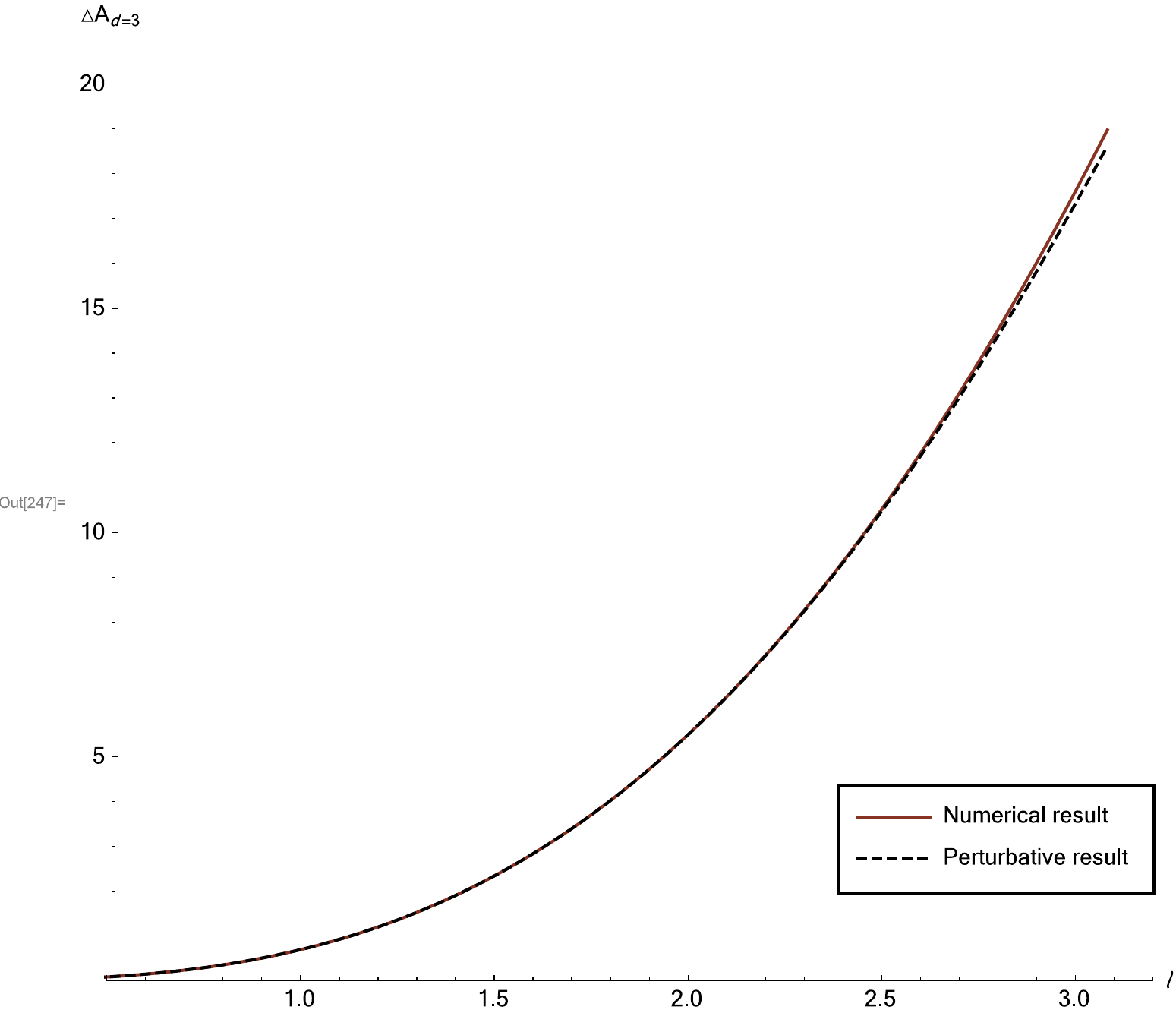}
		\caption{}
		\label{fig:NEE4D}
	\end{subfigure}
	\caption{Comparison of perturbative and numerical results for HEE in (a) $(2+1)$ and (b) $(3+1)$-dimensions, plots generated for $L=1$ and $z_h = \sqrt{10}$, the dashed curves are obtained from perturbation series.}
	\label{fig:NEE34}
\end{figure}

\begin{figure}
	\begin{subfigure}{0.49\textwidth}
		\centering
		\includegraphics[width=\textwidth]{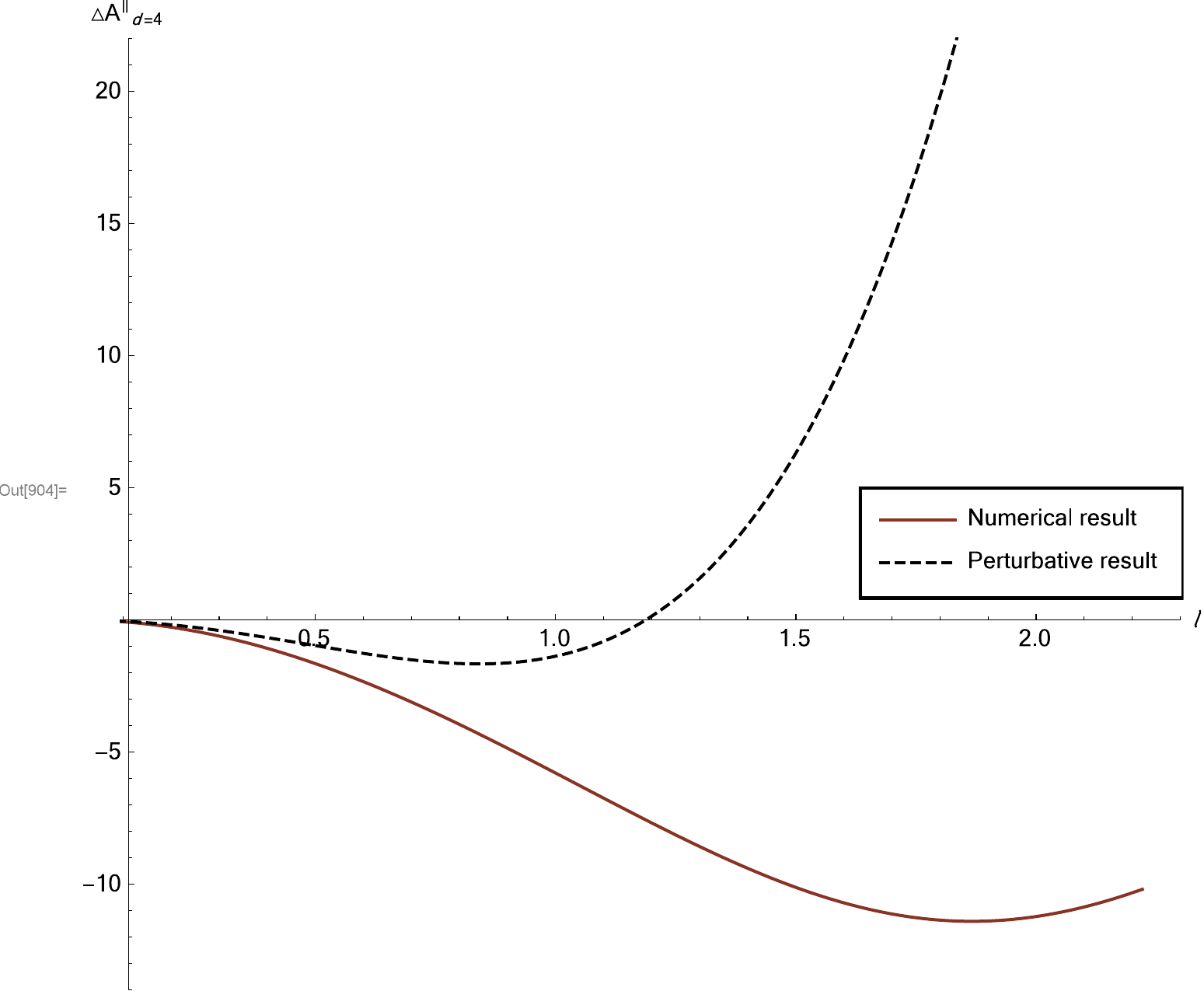}
		\caption{Strip parallel to the electric field.}
		\label{fig:NEE5D1}
	\end{subfigure}
	\hfill
	\begin{subfigure}{0.49\textwidth}
		\centering
		\includegraphics[width=\textwidth]{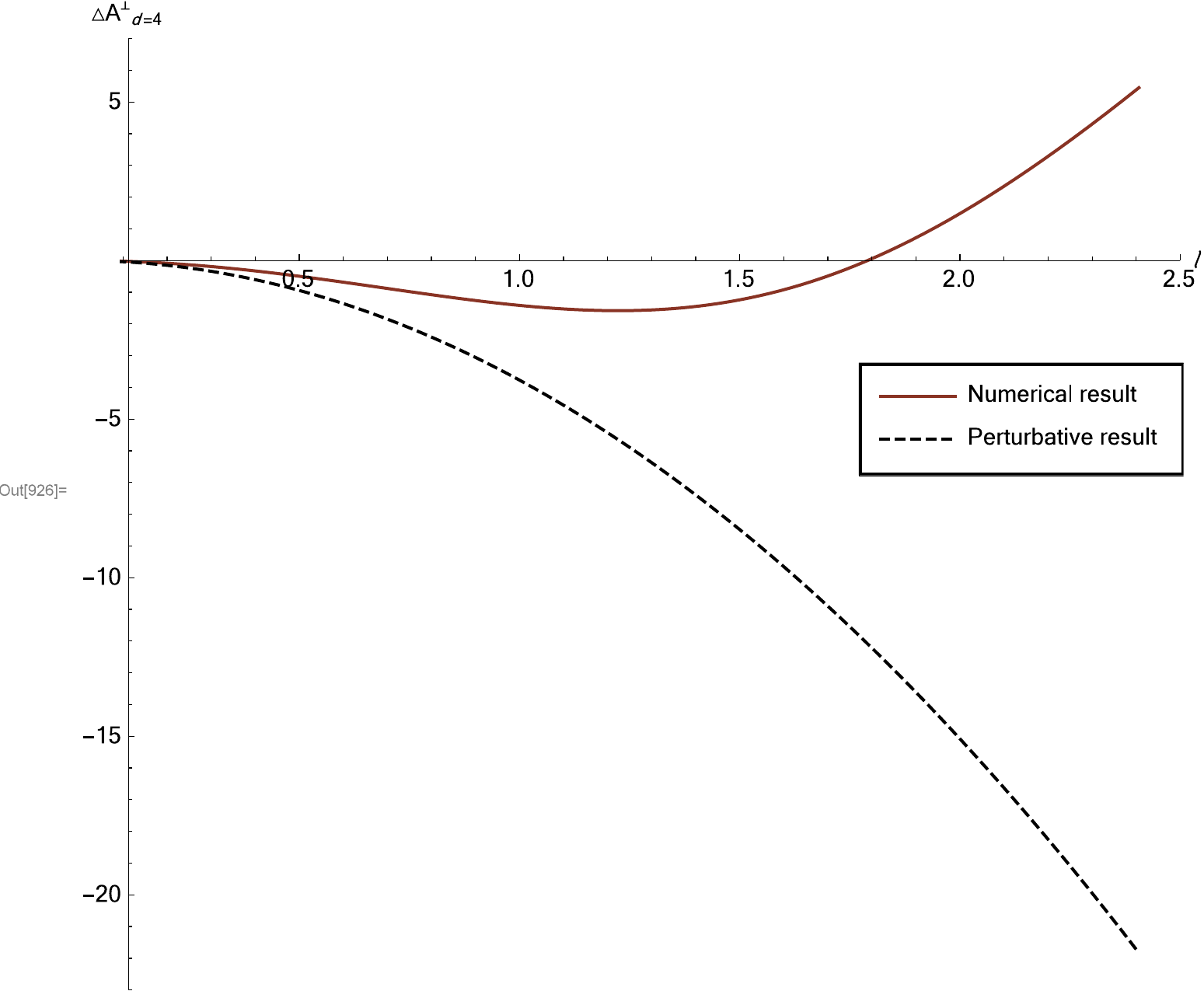}
		\caption{Strip perpendicular to the electric field.}
		\label{fig:NEE5D2}
	\end{subfigure}
	\caption{Comparison of perturbative and numerical results for HEE in $(4+1)$ dimensions, plots generated for $L=1$ and $z_h = \sqrt{10}$, the dashed curves are obtained from perturbation series.}
	\label{fig:NEE5D}
\end{figure}

 The results are summarized in figures \ref{fig:NEE34} and \ref{fig:NEE5D}. It is evident that the perturbative results mimic the actual ones very closely in three and four dimensions and they begin to differ as $l \to z_h$. From Figure (\ref{fig:NEE5D}) we also note that the HEE does start from negative values for both parallel and perpendicular cases in five dimensions as predicted by our approximate calculations. However, in this case the perturbative results do not seem to replicate the actual results as closely as in the other two dimensions, demanding to move higher in the perturbation order.


\subsection{Holographic Sub-region Complexity:}
 In the following, we calculate the volume $(V_{\gamma})$ and estimate $C_A$ using \eqref{hsc} within the same perturbative framework used for entanglement entropy. The calculations are usually cumbersome and hence, we restrict ourselves to first non-zero order only.
\par Let us choose the $(3+1)$ dimensional osm \eqref{Ads4} for illustration. The volume of a co-dimension $1$ minimal hypersurface homologous to the boundary subregion may be found by solving the double integral
\begin{align}\label{Volume4}
    V_{\gamma} &= 2 L\int_{\delta}^{z_*}
    \frac{dz}{z^3\sqrt{1-E^2z^4}}\int_{0}^{x(z)}dx ~,\nonumber \\
    &= 2 L\int_{\delta}^{z_*}
    \frac{dz}{z^3\sqrt{1-E^2z^4}}\int_{z}^{z_*} \frac{\frac{\tilde{z}^2}{z_*^2}d\tilde{z}}{\sqrt{(1-\frac{\tilde{z}^4}{z_*^4})(1-E^2\tilde{z}^4)}}~.
\end{align}
As before, we can expand the denominators in a power series of $E^2z_*^4$ and then solve the integrals in terms of beta functions. One can show with little effort that,
\begin{equation*}
    V_{\gamma} = V_{(0)} -\frac{3\pi}{320}  \frac{ E^2 L l^3}{b_0^4}~,
\end{equation*}
where $V_{(0)}$ is the volume in pure $AdS_4$. Thus the leading non-zero correction occurs at $\mathcal{O}(E^2)$ and estimated to be
\begin{equation}\label{scomconf3}
    \Delta C_{A(4)} = -\frac{3}{320}  \frac{E^2 L}{8G_{eff}^{(4)}} \frac{l^3}{b_0^4}~.
\end{equation}
Similar calculations in three and five dimensions lead to
\begin{align}
    \Delta C_{A(3)} &= \frac{E^2}{8G_{eff}^{(3)}}\frac{l^4}{64}~,~~~\nonumber \\
    \Delta C_{A(5)} &= \frac{E^2 L^2}{8G_{eff}^{(5)}}\frac{l^2}{2b_0^2}\left[\frac{\pi}{54b_0} + \frac{\pi I_1}{18b_0^2} -\frac{7}{108} B\left(\frac{5}{6}, \frac{1}{2}\right) \right]~~~~\text{parallel~ strip}~, \nonumber \\
    &= \frac{E^2 L^2}{8G_{eff}^{(5)}}\frac{l^2}{2b_0^2}\left[\frac{\pi}{54b_0} + \frac{\pi \tilde{I_1}}{18b_0^2} -\frac{5}{54} B\left(\frac{5}{6}, \frac{1}{2}\right) \right]~~~~~\text{orthogonal~strip}~.
\end{align}
The coefficients are written in terms of relevant beta integrals defined earlier. Note that, at leading order the change in complexity in four dimensions is negative whereas in other dimensions it is positive. We will get back to these later during the discussions.

\subsection{HEE in the high temperature regime: }
Let us recall the relevant integrals for the size of the subsystem and the area of RT surface in three dimensions,
\begin{eqnarray}
    \frac{l}{2} &=& z_*\int_{0}^{1}\frac{y~\sqrt{1+E z_{*}^2}~ dy}{\sqrt{1-y^2}\sqrt{1-E^2z_*^4 y^4}}~, \label{27} \\
    \mathcal{A} &=& 2\int_{\epsilon}^{1}\frac{dy}{y}\sqrt{\frac{1+E z_*^2y^2}{(1-E z_*^2 y^2)(1 -y^2)}}~. \label{28}
\end{eqnarray}
We already know that the geometry exhibits an effective Hawking temperature $T_{eff} \sim \sqrt{E}$, while the turning point is given by $z_{*} \sim l$; therefore, $Ez_{*}^2 \sim l^2T_{eff}^2$. We have been working exclusively in the low effective temperature regime $E\rightarrow 0$ such that $Ez_{*}^2 \ll 1$. It is worthwhile to check the behaviour of entanglement entropy in the high temperature regime as well, where $Ez_{*}^2 \to 1$ $(z_*\rightarrow z_h)$. In fact, we expect the entanglement entropy to match exactly with the Bekenstein-Hawking entropy as shown in earlier works \cite{Bhattacharya:2017gzt, Chakraborty:2014lfa}.
\par The idea is to replace the integral in \eqref{28} with the one in \eqref{27}, because in the $Ez_{*}^2 \to 1$ limit both of them are dominated by the poles at $y=1$. This leads to the simple expression
\begin{equation}
    \mathcal{A} \simeq  \frac{\sqrt{2} l}{z_*} = \sqrt{2} ~l\sqrt{E}~,
\end{equation}
and subsequently
\begin{equation}
    S_{E} \simeq \sqrt{2}\frac{l\sqrt{E}}{4G_{eff}^{(3)}}~.
\end{equation}
The Bekenstein-Hawking entropy can be found by setting $z = z_h$ and calculating area of the horizon, giving
\begin{equation}
    S_{BH} = \sqrt{2}\frac{l\sqrt{E}}{4G_{eff}^{(3)}}~.\label{sbh}
\end{equation}
Thus the HEE matches precisely with the B-H entropy at high temperature. Furthermore, there is another interesting observation that we want to stress upon at this point. It can be easily checked that, for a non-rotating BTZ black hole with mass parameter $m$, the B-H entropy is precisely given by \eqref{sbh}, provided we set the BTZ at the exact same temperature \eqref{osmtemp}. Again for BTZ, we know that at high temperature, the HEE merges with the B-H entropy \cite{Swingle:2011np, Dong:2012se,Ryu:2006bv,Bhattacharya:2017gzt}. From this chain of arguements, we conclude that in the high temperature regime, the HEE in $AdS_3$-osm matches exactly with that in BTZ. So, even though there are deviations at various orders in the perturbative regime \eqref{osmbtzst}, the differences are washed away at high temperatures.\\
Finally, similar calculations in rest of the cases yield:
\begin{center}
\begin{tabular}{|c|c|c|}
    \hline
     $d$ & $S_E$ & $S_{BH}$  \\
     \hline
     $3$ & $\frac{l L E}{4G_{eff}^{(4)}}$ & $\frac{l L E}{4G_{eff}^{(4)}}$\\
     $4$ & $\frac{2}{3}\frac{l L^2 E^{\frac{3}{2}}}{4G_{eff}^{(5)}}$ & $\frac{2}{3}\frac{l L^2 E^{\frac{3}{2}}}{4G_{eff}^{(5)}}$\\
     \hline
\end{tabular}
\end{center}

The exact matching of the Bekenstein-Hawking entropy and high temperature limit of entanglement entropy for these open string geometries is a very exciting result. In fact, this strengthens our intuition that even for these kinematic spacetimes, there exists an intricate relationship between entanglement (information) and geometry. Although it is quite well understood within Einstein gravity, this is something new for the open string geometries where the horizon is \textit{induced} by introducing electric fields. Put in another way, the introduction of electric field gives rise to some kind of entanglement that results in the formation of these black hole kind of geometries which, in spite of being non-Einsteinian, carries a signature of the entanglement, manifest in this matching in all the dimensions studied in this paper. This also attributes a physical significance to the area of these emergent horizons in the boundary theory, thus resolving  the ambiguity raised in \cite{Kundu:2013eba,Kundu:2015qda,Banerjee:2015cvy} where the authors failed to identify it with the thermal entropy.
%%%%%%%%%%%%%%%%%%%%%%%%%%%%%%%%%%%%%%%
\section{HEE and HSC for spherical subregion } \label{sec4}
%%%%%%%%%%%%%%%%%%%%%%%%%%%%%%%%%%%

In this section, we study the changes in HEE and HSC for spherical entangling region, upto orders similar to the strip case. In general the two choices for the entangling region correspond to two different ways of choosing a mixed state. They also lead to different bipartition or factorization of the boundary Hilbert space and the idea is to figure out the quantities that are insensitive to the choice of the entangling region. The spherical subregion choice is made by defining $\sum_{i=1}^{d} x_{i}^{2} = R^{2}$, where $R$ is the radius of the region. Note that, in this case, the change in area and volume will be related to the  change in HEE and HSC as \cite{PhysRevD.100.126004} ,
\begin{equation}
    \Delta S^{sph}=\frac{\Omega_{d-2} \Delta \mathcal{A}}{4 G_{eff}}~,
\end{equation}
\begin{equation}
    \Delta C^{sph}_{A}=\frac{\Omega_{d-2} \Delta \mathcal{V}}{8 \pi G_{eff} \left( d-1\right)} ~,
\end{equation}
where $\Omega_{d-2}$ is $(d-2)$ dimensional volume of unit radius $S^{(d-2)}$. The normalization has been chosen for convenience.

Note that, in (2+1) dimensions the choices for the two subregions are related to each other  since in that case, the number of boundary spatial dimensions is simply one, which becomes either the small strip length or the radius of the sphere. The results for the spherical subregion thus trivially follow from those for the strip region upon the identification $l= 2R$, where $R$ is the radius of the sphere. Hence we refrain ourselves from presenting the details of calculation in (2+1) dimensions and directly move to higher dimensions.

\subsection{(3+1) dimensions :}
  With the choice of spherical subregion, the $(3+1)$ dimensional osm can be written in the following form,
 \begin{equation}
     ds_{(4)}^{2} = \frac{1}{u^{2}} \left[-\left( 1- E^{2} u^{4} \right) dt^{2} + \left(\frac{1}{1- E^{2} u^{4}} \right) du^{2} +  dr^{2} + r^{2} d\theta^{2} \right] ~.
 \end{equation}
In this case, the sensible choice of perturbation parameter is $\lambda= E^{2} R^{4}$. Now following \cite{PhysRevD.100.126004}, we make the following reparametrizations
\begin{equation}
    z(x)=\frac{u(r)}{R} \, \,   , \, \, x=\frac{r}{R}~. \label{rep}
\end{equation}
under which the metric on the codimension-$2$ surface takes the following form,
 \begin{equation}
     ds^{2} = \frac{1}{z(x)^{2}} \left[  \left(\left( 1+ \lambda z(x)^{4} \right) +   \left(\frac{1}{1-\lambda z(x)^{4}} \right) z'(x)^2 \right) dx^{2} + x^{2} d\theta^{2} \right] ~.
 \end{equation}
 Then the area and the volume integral in this case are given by,
 \begin{eqnarray}
     \mathcal{A} &=& \int d\theta \int dx \frac{x}{z(x)^{2}} \left[\left( 1+ \lambda z(x)^{4} \right) +   \left(\frac{1}{1-\lambda z(x)^{4}} \right) z'(x)^2 \right]^{\frac{1}{2}}~,\\
      \mathcal{V} &=& \int_{0}^{2\pi} d\theta \int_{0}^{1} dx~ x \left(\int_{\epsilon}^{z(x)} \frac{du}{u^3 \sqrt{1-\lambda u^4}}\right)~.
 \end{eqnarray}
Next we solve for the Euler-Lagrange equation resulting from the extremization of the area integral. With an embedding of the form 
\begin{equation}
    z(x) = \sqrt{1-x^2} + \lambda z_{1}(x)~, \label{fembd}
\end{equation}
the solution turns out to be
\begin{equation}
     z^{(1)}(x) = \frac{3 x^6-11 x^4+17 x^2+8 \sqrt{1-x^2}-8 \log \left(1+\sqrt{1-x^2}\right)-9}{30 \sqrt{1-x^2}}~.
 \end{equation}
 Using this, the change in area upto second order is given by,
 \begin{equation}
    \Delta \mathcal{A}^{(1)} = \frac{\lambda}{15}~,~~~~ ~~~~~~~\Delta \mathcal{A}^{(2)} = -\frac{1680 \log (2)-1033}{23625} ~\lambda^{2}~.
 \end{equation}
 Hence, the change in HEE can be written in the following form,
 \begin{equation}
     \Delta S^{sph}_{(4)} = \frac{E^2 R^4}{4 G_{eff}^{(4)}}\left[\frac{2 \pi}{15} - \frac{2 \pi\left( 1680 \log(2)-1033\right)}{23625} E^2 R^4 \right] +\mathcal{O}\left(E^6\right)~.
 \end{equation}
 But the surprising result in this case is that the first order change in HSC, similar to the strip case, turns out to be non-zero, given by
 \begin{equation}
   \Delta C_{A(4)}^{sph} = -\frac{E^2 R^4}{128  G_{eff}^{(4)}}~.
 \end{equation}
 In all the studies carried out so far with  various kind of geometries, the first order change in volume for spherical subregion is always found to be zero. Thus far, this turns out to be a distinct and remarkable feature of the open string geometry and we will get back to it in the discussions.

\subsection{(4+1) dimensions:}
The five-dimensional osm is given by
\begin{equation}
    ds_{(5)}^{2} =\frac{1}{u^2} \left[- \left(1- E^{2} u^{4}\right) dt^{2} + \frac{du^{2}}{\left(1-E^{3}u^{6}\right)}+\frac{1+E u^{2}}{\left(1+ E u^{2} + E^{2} u^{4}\right)}dx_{||}^{2} + dx_{\perp}^{2}\right]~.
\end{equation}
 As the spatial part of the boundary is three dimensional, we will switch to the spherical polar coordinates and parametrize the boundary spatial directions in the following way
\begin{equation}
    x_{|| } =r \sin{\phi} \cos{\theta} ~,~~~ x_{\perp 1} = r \cos{\phi}~ ,~~~~ x_{\perp 2} = r \sin{\phi} \sin{\theta}~.
\end{equation}
In terms of these spherical polar coordinates, the constant timeslice of the metric is not quite illuminating. We further make the following reparametrization

%\begin{equation}
%\begin{align}
    %ds_{i}^{2} = \frac{1}{u^{2}} \left[ \frac{du^{2}}{1 - E^3 u^{6}} &+ \frac{1}{2} \left(\sin^{2}{\phi}  ((p_1-1) \cos{2 \theta}+p_1+1)+2 \cos^{2}{\phi }\right) dr^{2} \notag
    %\\
    %&\quad + r^2 \left(\cos^{2}{\phi} \left(\sin^{2}{\theta}+p_1 \cos^{2}{\theta }\right)+\sin^{2}{\phi }\right) d\phi^{2} \notag \\ 
    %&\qquad\qquad +\frac{1}{2} r^2 \sin^{2}{\phi } (-(p_1-1) \cos{2 \theta}+p_1+1) ~d\theta^{2} \notag \\
    %&\qquad\qquad -(p_1-1) r \sin{2 \theta} \sin^{2}{\phi } dr d\theta +(p_1-1) r \cos^{2}{\theta } \sin{2\phi}d\phi dr \notag 
    %\\
    
    %&\qquad\qquad  - \frac{1}{2}(p_1-1) r^2 \sin{2\theta }  \sin{2\phi ) d\phi d\theta \right].  
%\end{align}
%\end{equation}
%\begin{equation}
%\begin{align}
%ds_{i}^{2} = \frac{1}{u^{2}} \left[ \frac{du^{2}}{1 - E^3 u^{6}} &+ \frac{1}{2} \left(\sin^{2}{\phi}  ((p_1-1) \cos{2 \theta}+p_1+1)+2 \cos^{2}{\phi }\right) dr^{2} \right.\\&+ r^2 \left(\cos^{2}{\phi} \left(\sin^{2}{\theta}+p_1 \cos^{2}{\theta }\right)+\sin^{2}{\phi }\right) d\phi^{2} \\ &+\frac{1}{2} r^2 \sin^{2}{\phi } (-(p_1-1) \cos{2 \theta}+p_1+1) ~d\theta^{2} \\& -(p_1-1) r \sin{2 \theta} \sin^{2}{\phi } dr d\theta +(p_1-1) r \cos^{2}{\theta } \sin{2\phi}d\phi dr \\&  - \frac{1}{2}(p_1-1) r^2 \sin{2\theta }  \sin{2\phi ) d\phi d\theta \right]
%\end{align}
%\end{equation}

%where $p_{1}=\frac{1+Eu^{2}}{1+Eu^{2}+E^2 u^{4}}$. 

\begin{equation}
    y=\frac{r}{R}~,~~ z= \frac{u}{R}~,~~ \lambda = E^2 R^4~,
\end{equation}
and choose an embedding of the form $z=z(y)$ . The area integral in this case takes the following form

\begin{equation}
    A =\int_{\theta=0}^{\pi} \int_{\phi=0}^{2\pi} \int_{y=0}^{1} d\theta d\phi ~dy ~\frac{f(y, \, \theta, \, \phi)}{z(y)^{3}}~,
\end{equation}
where\\
$
\begin{footnotesize}
f(y, \,\theta, \, \phi)= y^2 \sin (\phi )\sqrt{\frac{  \left(- z^{\prime}(y)^2 \left(\lambda  z(y)^{4} \cos ^2(\theta ) \sin ^2(\phi )+\sqrt{\lambda } z(y)^{2}+1\right)+\sqrt{\lambda } z(y)^{2} \left(\lambda ^{\frac{3}{2}} z(y)^{6}+\lambda  z(y)^{4}-1\right)-1\right)}{\left(\sqrt{\lambda } z(y)^{2}-1\right) \left(\lambda  z(y)^{4}+\sqrt{\lambda } z(y)^2+1\right)^2}}.
\end{footnotesize}
$
Repeating the same algorithm, we arrive at the following change \footnote{The presence of the fractional $3/2$-th order term might be surprising. But this arises due to the typical nature of the metric with a $E^{3} u^{6}$ term and the way we have defined the perturbation parameter $\lambda=E^{2} R^{4}$. However this is quite similar to the standard $AdS_{5}$ BH case where $E^4$ should be the $2$nd order. Therefore, we write $E^3$ order as $3/2$-th order.} 
\begin{equation}
    \Delta \mathcal{A}^{(1)} = - \frac{8 \pi}{15} \lambda   ~ , \,~~ \Delta \mathcal{A}^{(\frac{3}{2})} =  \frac{12 \pi}{35} \lambda^{\frac{3}{2}} ~ , ~~ \Delta \mathcal{A}^{(2)} = - \frac{1376 \pi}{23625} \lambda^{2}~.
\end{equation}
Recast in terms of $E$, the change in HEE is
\begin{equation}
    \Delta S_{(5)}^{sph} = \frac{E^2 R^4}{4G_{eff}^{(5)}}\left[-\frac{32  \pi^{2} }{15} + \frac{48\pi^2}{35} E R^2\right] + \mathcal{O}(E^4)~. \label{hee5}
\end{equation}
The volume integral in this case is 
\begin{equation}
   \mathcal{V}= \int_{\theta=0}^{\pi} \int_{\phi=0}^{2\pi} \int_{y=0}^{1} d\theta d\phi dy y^{2}\left(\int_{\epsilon}^{z(y)} \frac{dp}{p^{4}}\sqrt{\frac{ \left(\sqrt{\lambda } p^2+1\right) \sin ^2(\phi )}{ \left(1-\sqrt{\lambda } p^2\right) \left(\lambda  p^4+\sqrt{\lambda } p^2+1\right)^2}}\right)~,
\end{equation}
and the results for the first order (in this case , we consider both $\lambda$ and $\lambda^{\frac{3}{2}}$ as first order terms in a way),
\begin{equation}
    \Delta \mathcal{V}^{(1)} = 0~, ~~~ \Delta \mathcal{V}^{(\frac{3}{2})} = - \frac{\pi^{2}}{8} \lambda^{\frac{3}{2}}~. 
\end{equation}
In this case, the change in HSC can therefore be written in the following form,
\begin{equation}
    \Delta C_{A(5)}^{sph} = - \frac{\pi^2 E^3 R^6 }{48 G_{eff}^{(5)}}+ \mathcal{O}\left(E^4\right)~.
\end{equation}
Note that, for the class of parametrization we are considering, i.e, the radial direction as function of a boundary spatial direction, there is a unique choice in case of spherical subsystem, irrespective of the direction of the electric field. So unlike the strip case, here we only have  one set of results for the change in HEE and HSC.

The leading order change in HEE is negative again like strip case, showing universality with respect to the choice of subregion. Unlike $AdS_{4}$-osm, the leading order change in volume in this case is zero, which is a bit satisfactory primarily. But, then one finds that at subleading order, the change is negative again.\footnote{The leading non-zero change in volume in AdS$_{4}$ and AdS$_{5}$-osm intrigues us to compute the same for $AdS_{3}$-osm. This however, turns out to be positive definite
\begin{equation*}
    \Delta C_{A(3)}^{sph} =  \frac{ E^2 R^4 }{32  G_{eff}^{(3)}}  ~.
\end{equation*}} Finally we conclude the section by recalling that in both four and five dimensions, where the leading non-zero change in HSC is negative, WEC condition is also violated.
%%%%%%%%%%%%%%%%%%%%%%%%%%%%%%%%%%%%%%
\section{Entanglement Thermodynamics upto first order :}\label{sec5}
%%%%%%%%%%%%%%%%%%%%%%%%%%%%%%%%%%%%%

In this section we will work out the entanglement thermodynamics in the probe sector, upto first order. Conventionally in holography, the stress tensor of the boundary field theory is read off from the subleading term in the Fefferman-Graham expansion of the asymptotically $AdS$ metric. Even though the open string metric is asymptotically $AdS$, there is no valid reason for Fefferman-Graham theorem to hold for these kinematic geometries, since they do not satisfy Einstein equations.  Hence, we will derive the  stress tensor due to the flavors in the boundary theory from the DBI action of the dual probe brane \cite{Banerjee:2015cvy, Karch:2008uy}. Towards that, first we define,

\be
U^{\mu \nu} = \frac{1}{\sqrt{-det \gamma}} \left[\frac{\delta S_{DBI}}{\delta \gamma_{\mu\nu}}+\frac{\delta S_{DBI}}{\delta \gamma_{\nu\mu}}\right]~,
\ee
where $\gamma_{\mu\nu} \equiv P\left[g_{\mu\nu}\right]$ is the induced metric on the brane worldvolume and we will choose the background metric  $g_{\mu\nu}$ to be that of pure $AdS_{d+1}$. For simplification, we will again consider space-filling brane embeddings so that  $P\left[g_{\mu\nu}\right]=g_{\mu\nu}$.
The boundary stress tensor is obtained by integrating out the directions along the brane which are transversal to the boundary. This gives

\begin{eqnarray}
\left\langle T^{a}_{b}\right\rangle &=& \int_{z_{min}}^{z_{max}} dz ~\sqrt{-det \gamma} ~U^a_b ~, \label{bdy}\\
&=& -N_f \tau_p  \int_{z_{min}}^{z_{max}} dz ~\left(det \gamma ~det \mathcal{S}\right)^{1/4} ~\mathcal{S}^a_b ~, \label{bdy2}
\end{eqnarray}
where the intermediate steps are detailed in appendix \ref{apenA}. Note that, in order to obtain boundary quantities like energy and pressure, we must integrate over the boundary coordinates, so the factor of $\sqrt{-det \gamma}$  in \eqref{bdy} ensures a covariant integral. Here $\ms^{\mu\nu}$ is the inverse of open string metric and $S^a_b= S^{a\mu}\gamma_{\mu b}$, where $a,b=0,...,d-1$ span the boundary directions.

Expectedly the integral \eqref{bdy2} is UV divergent, demanding regularization. Following the standard holographic prescription, we will introduce a UV cut-off at $z_{min}= \epsilon$, add covariant counterterms on the radial-slice and at the end send $\epsilon \rightarrow 0$.  Typically, the counterterm that we will add is 

\be
\mathcal{L}_{ct} = \frac{N_f \tau_p}{d} \sqrt{-det ~h_{ab}} ~,\label{cterm}
\ee
where $h_{ab}$ is the induced metric on the slice. Now, we can obtain this induced metric by pulling back either  the closed string geometry $g_{\mu\nu}$ or the open string geometry $\ms_{\mu\nu}$. In either of the cases, the counterterm cancels the divergence, but the finite contributions they yield differ, leading to a degeneracy in the regularization scheme. For our purpose, we will simply pullback the closed string geometry and not dwell upon this ambiguity any further. There is also a subtlety in the choice of the upper limit $z_{max}$. Typically for usual thermodynamic computations \cite{Banerjee:2015cvy}, it is chosen to be the open string horizon $z_h$. However, for the purpose of entanglement thermodynamics we will choose $z_{max}= z_*$, the turning point of the RT surface in the bulk.

To illustrate, we consider the open string geometry in four dimensions. In this case  $\gamma_{\mu \nu}$ is that of pure $AdS_4$, whereas $\ms_{\mu\nu}$ is given by  \eqref{osm4} . Then from \eqref{bdy2} we get,

\bea
\left\langle T^{t}_{t}\right\rangle &=&  \lim_{\epsilon\to 0} N_f \tau_p \left[\frac{1}{3 z^3}- E^2 z\right]_{\epsilon}^{z_*}=  \lim_{\epsilon\to 0}  N_f \tau_p\left(\frac{1}{3 z_*^3}-\frac{1}{3 \epsilon^3}- E^2 z_*\right)~,\\
\left\langle T^{x}_{x}\right\rangle &=& \frac{N_f \tau_p}{3}\left(\frac{1}{ z_*^3}-\frac{1}{ \epsilon^3}\right)=\left\langle T^{y}_{y}\right\rangle~.
\eea
Now for $d=3$ the renormalized stress tensor is given by
\be
\left\langle T^{a}_{b}\right\rangle _{ren}= \left\langle T^{a}_{b}\right\rangle + \frac{N_f \tau_p}{3} \sqrt{-det ~h} ~\delta^a_b = \left\langle T^{a}_{b}\right\rangle + \lim_{\epsilon \rightarrow 0} \frac{N_f \tau_p}{3 \epsilon^3}  ~\delta^a_b~.
\ee
This yields 
\bea
\left\langle T^{t}_{t}\right\rangle_{ren}&=&N_f \tau_p \left(\frac{1}{3 z_*^3}- E^2 z_*\right) ~,\label {ttren}\\
\left\langle T^{x}_{x}\right\rangle_{ren}&=& \frac{N_f \tau_p}{3 z_*^3}= \left\langle T^{y}_{y}\right\rangle_{ren}~.\label{xxren}
\label{norml}
\eea
Note that the first term in (\ref{ttren}) and the only term in (\ref{xxren}) correspond to pure $AdS_4$ results. So the change in energy and pressure is given by is given by
\be
\Delta \xi = \int dx dy \left\langle T_{tt}\right\rangle_{ren} =  N_f \tau_p E^2 z_* l L= \frac { N_f \tau_p E^2 l^2 L}{2b_0} + \mathcal{O}(E^4)~,~~~~\Delta P_x = 0~,
\ee
where we have used that fact that at leading order $z_* \sim \frac{l}{2b_0}$. Now from (\ref{ads4}) the first order change in entanglement entropy is given by 
\be
\Delta S_{(4)} = \frac{E^2 l^3 L }{80 ~G_{eff}^{(4)} ~b_0^2} + \mathcal{O}(E^4)~.
\ee
So the entanglement temperature is given by 
\be
T_{ent}= \frac{\Delta \xi - \frac{d-1}{d+1} V_{d-1}\Delta P_x}{\Delta S_{(4)}} =\frac{40 ~G_{eff}^{(4)} ~b_0~ N_f \tau_p}{l} \sim \frac{1}{z_*}~.
\ee
%%%{ Next we check whether in the large temperature limit the first law of entanglement boils down to the first law of black hole thermodynamics. In this case, $z_* \rightarrow z_h$, so from \eqref{norml} we have  $\left\langle T^{t}_{t}\right\rangle_{ren}$\footnote{Note that in case of pure $AdS_4$, $z_h \rightarrow \infty$, so \ref{norml} vanishes in the lower limit. So in presence of electric field, both the terms contribute to the shift in the stress tensor.} $= -\frac{ 2 }{3}N_f \tau_p E^{3/2} $. So the change in energy is given by
%%\be
%%%%\ee
%%\sqrt{E} l \rightarrow \mathcal{O}(1)$. Similarly, the change in entanglement entropy is $\Delta S \sim E^2 l^3 L \sim \sqrt{E} L$. Hence
%%\be
%%T_{ent}^{high} \simeq \frac{\Delta E}{\Delta S} \sim E^{1/2} \sim T_{eff}
%%\ee
Next we turn our attention to three and five dimensional cases, where there are log-terms in the flavor stress tensor\footnote{Recall that Fefferman-Graham expansion in even boundary dimensions also admits a log term.}. The regularization of the log-divergence leads to introduction of an arbitrary scale in the system and hence logarithmic violation of conformal invariance.  To illustrate upon the consequence of the log-term, let us consider the three-dimensional case. In this case, components of the stress tensor turn out to be
\begin{eqnarray}
\langle T^{t}_{t}\rangle &=& \lim_{\epsilon\to 0} ~N_{f} \tau_{p} \left[\left(\frac{1}{2z^2}- E\right){\sqrt{1+E z^2}} + \frac{1}{2} \log \left(\frac{z}{1+\sqrt{1+ E z^2}}\right)\right]_{\epsilon}^{z_*}, \\
\langle T^{x}_{x}\rangle &=& \lim_{\epsilon\to 0} N_f \tau_p\left[\frac{1}{2z^2}\left\{\sqrt{1+Ez^2}+ E z^2~ \log \left(\frac{z}{1+\sqrt{1+ E z^2}}\right)\right\}\right]_{\epsilon}^{z_*}.
\end{eqnarray}
 The substitution of the limits with the assumption $Ez_*^2 \rightarrow 0$ results in
\bea
\left\langle T^{t}_{t}\right\rangle &=& \lim_{\epsilon\to 0} ~N_f \tau_p\left[\left(\frac{1}{2z_*^2}-\frac{1}{2\epsilon^2}\right) - \frac{E}{4}\left(1 + 2 \text{ log}\left(\frac{z_*}{\epsilon}\right)\right)\right]~,\\
\left\langle T^{x}_{x}\right\rangle &=&\lim_{\epsilon\to 0} ~N_f \tau_p\left[\left(\frac{1}{2z_*^2}-\frac{1}{2\epsilon^2}\right) + \frac{E}{4}\left(-1 + 2 \text{ log}\left(\frac{z_*}{\epsilon}\right)\right)\right]~.
\eea
So now there is log-divergence in addition to the usual $1/\epsilon^2$ term. The regularization of such stress tensor has been discussed in details in \cite{Banerjee:2015cvy}. The choice of suitable counterterm along with  (\ref{cterm}) gives
\be
\Delta \xi = \frac{N_f \tau_p E l}{4}~\left[1 + 2\text{log}\left(\frac{z_*}{z_0}\right) \right]~,~~~~~~\Delta P_x = \frac{N_f \tau_p E }{4}~\left[-1 + 2\text{log}\left(\frac{z_*}{z_0}\right)\right]~.
\ee
Therefore
\be
\Delta \xi  - \frac{d-1}{d+1} ~l \Delta P_x = \frac{N_f \tau_p E l}{3}\left[1+ \text{log}\left(\frac{z_*}{z_0}\right)\right]~.
\ee
The log-term is just the reminiscent of breaking of conformal invariance. Ignoring it and using (\ref{hee3}) upto leading order, the entanglement temperature is given by
\be
T_{ent} = \frac{16 G_{eff}^{(3)}N_f \tau_p}{3 l}~.
\ee
These results are quite significant. Conventionally the stress-tensor (hence energy and pressure)  of the boundary theory is computed from the Fefferman-Graham expansion of the metric, which as we pointed out earlier, is supposedly not a valid expansion for non-Einstein geometries. However, the stress tensor computed from the DBI action does the job for us in this case, giving rise to the desired form of thermodynamic relation in the perturbative regime. Note that at this point, it is quite tempting to take the high temperature limit, $z_* \rightarrow z_h$ of the entanglement temperature which will relate it to the Hawking temperature $T_{eff}$. However, as pointed out earlier, the thermal entropy of the flavor sector, which in our kinematic setup is in a NESS, is not well-defined. So at the level of the first law of thermodynamics, this high temperature limit is ambiguous and we refrain from further commenting on that.

\section{Results and Discussion} \label{sec6}
Intrigued by the wide range of similarities open string geometries share with black hole solutions of Einstein gravity, we have explored various aspects of entanglement entropy and complexity in the former, using the standard holographic prescriptions implemented in the latter. Our study is mostly perturbative where we treat these geometries as excitations over empty $AdS$, with the electric field playing the role of the perturbation parameter.  Let us summarize the results in various dimensions: \vspace{1mm}

In three dimensions, the results are most satisfactory. The changes in HEE upto subleading order show up with plausible signs for both strip and spherical entangling region and their numerical values are precisely the same, which is trivially expected in (2+1) dimensions. Furthermore, we went on to show that in the perturbative regime, $AdS_3$-osm is more entangled compared to a BTZ black hole at the same temperature and the degree of entanglement increases as we move up in orders of perturbation. This difference however disappears in the high-temperature regime and the two results match exactly. The change in HSC at leading order vanishes for both choices of subsystem and this is again in perfect agreement with BTZ results. \vspace{1mm} 

Peculiarities tend to show up as we move higher in dimensions. In four dimensions, the leading order change in HEE is positive for both the choices of subsystem whereas the leading order change in HSC is negative in either of the cases. The five dimensional case is sufficiently involved and comes with more variety of features. Firstly, in five dimensions there is an anisotropy in the boundary spatial directions induced by the electric field, leading to two distinct choices for the strip region: shorter length along the electric field and perpendicular to it. However, in either of the cases, the leading order change in HEE turns out to be negative. For spherical subsystem, there is a unique choice for embedding, but similar conclusion holds as well. Thus in five dimensions, the excited state induced by the electric field is less entangled compared to the ground state, at least at leading order. However, as the numerical plots in Figure (\ref{fig:NEE5D})  suggest, if we consider full non-perturbative effects, then beyond certain $l$, say $l_*$, change in HEE tends to become positive.  This critical value $l_*$ becomes smaller as we keep increasing the value of  the electric field. Finally, the leading order results for change in HSC in this case turn out to be plausible for either choices of the entangling region. So to summarize, for higher dimensional osms, departures from familiar results show up either in HEE or in HSC depending on the dimensions. This in turn reflect upon that fact that these finite-temperature NESS are excited rather unconventionally by turning on electric field as compared to usual temperature deformations. These results are quite unfamiliar to Einstein gravity, but at the same time, if we try to view these higher dimensional geometries within Einstein gravity, they violate the WEC. 

Apart from these couple of peculiar features, there are some remarkable results as well. First of all, we have shown that in the high temperature regime, the entanglement entropy for the choice of strip entangling region, is in perfect agreement with the Bekenstein-Hawking entropy associated with the osm event horizon. This gives further support to the identification of the horizon area in the boundary theory and resolves the previous mystery regarding its mismatch with the thermal entropy, which is nevertheless an ill-defined concept for a system in NESS. The exact matching in all the three dimensions we consider in this paper only strengthens the deep connection between geometry and entanglement, which seems to hold even for these emergent geometries and the NESS. This is an interesting observation deserving further exploration. \vspace{1mm}

 We have also shown the validity of the first law of entanglement thermodynamics at leading order for strip entangling region. Our results are significant in the sense that we have derived the energy and pressure in the boundary theory from the DBI action of the probe brane, bypassing the questionable validity of Fefferman-Graham theorem for non-Einstein geometries. However, this still results in well-behaved entanglement temperature with smooth high temperature limit. Lastly, we have also realized during our analysis and also briefly elaborated in  Appendix \ref{apenB} that the embedding function theorem holds good in these open string geometries. \vspace{1mm}

There is a vital leap of faith that we make in this study. We assume that for applying the holographic Ryu-Takayanagi proposal, it is sufficient to work with an asymptotically $AdS$ metric, which the open string metrics are, irrespective of the origin. This might be a questionable assumption, but we would like to take the creative liberty in this case exploiting the lack of clarity on this topic. The aspects of entanglement in gauge theories and gravity are yet to reach a complete understanding and we believe it is too early to completely discard the study of entanglement for non-Einstein solutions. Interestingly, some of our results turn out to be in excellent agreements with those of Einstein gravity, going in favour of our assumption. But expectedly there are certain departures, especially in higher dimensions, which nevertheless makes our study meaningful. We attribute these departures to the pathologies of these geometries with respect to the energy conditions and the fact that the states we talk about in the boundary are in a strict sense in  NESS rather than being in perfect equilibrium. It will be indeed interesting to explore these connections further which may lead to new insights into our understanding of entanglement, complexity and the robustness of the RT proposal. This however, we will leave for future studies. 

There are a few other interesting avenues that can be explored continuing along this line of study. Throughout the work we have mostly restricted ourselves to the perturbative sector, within which it seems impossible to address physically interesting scenarios such as backreaction of the flavor on the gluon bath. However, with the recent progress in numerical holography in the context of entanglement \cite{Ecker:2015kna, Ecker:2018jgh}, this can certainly be explored. Also our setup is qualitatively very similar to the ones studied in \cite{Narayan:2012ks,Narayan:2013qga,Mukherjee:2014gia}. Following them, it will be interesting to study phase transition and mutual information within out setup. The introduction of electric field is also known to give rise to non-commutative effects \cite{Seiberg:2000ms} and there exists a finite non-locality scale \cite{Berkooz:1998st} for field theories in such non-commutative backgrounds. Our setup is perfectly equipped to capture the imprints of such non-locality in entanglement entropy, if any.

Despite the exact matching between the entanglement entropy and Bekenstein-Hawking entropy across all the dimensions, one might wonder about the validity of Ryu-Takayanagi prescription in open string geometries which are non-dynamical. To answer this unambiguously, one needs to explore the generalized gravitational entropy prescription put forward in \cite{Lewkowycz:2013nqa,Faulkner:2013ana,Huang:2014aga} in the present context.  Towards that, we propose a very simple setup. One can easily check that  the generalized gravitational entropy of a scalar field in the (2+1) dimensional open string metric precisely matches with that in BTZ as discussed in \cite{Lewkowycz:2013nqa}, provided that the two geometries are set to be at the same temperature, which can be achieved by adjusting the electric field. Now to check whether this gives the area law one needs to calculate the change in the area of the osm horizon. Now the open string metric being non-dynamical, this is hard to capture directly. Nevertheless, this can be circumvented as follows. One can think of the scalar as a worldvolume scalar of the probe branes and consider a stack of such branes. Now we can single out a brane and consider the backreaction of the rest on the background geometry which satisfies Einstein's equation.  The resulting change in the background metric changes the open string metric according to  (\ref{osm})  and then one can compare between the changes in osm horizon  area and generalized entropy for the scalar. This is part of an ongoing work which will appear soon. Finally, the plausibility of results in $AdS_3$-osm inspires us to go beyond RT proposal and explore more general aspects like time-dependent entanglement entropy \cite{Hubeny:2007xt} in the open string geometries, which we again leave for future investigations.

\section*{Acknowledgements}
AB$^{(1,2)}$ would like to thank Shibaji Roy and Arnab Kundu for several useful discussions.  We also thank Justin David for encouragement at the early stages of the work. We particularly thank K. Narayan for constructive feedback as well as bringing various relevant references into notice. Finally the authors would like to thank the people and the government of India for making research in theoretical HEP possible even at the most testing times the country has ever  faced.
\appendix
\section{Variation of DBI action:} \label{apenA}
Schematically the DBI action is given by 

\be
S_{DBI} = -N_f \tau_p \int d^{p+1}y \sqrt{-det M_{ab}}~ , ~~~~~~~~~~M_{ab}= \gamma_{ab}+ F_{ab}~,
\ee
where we have set $2\pi \alpha'=1$. Now the variation of the action is given by

\bea
\nonumber \delta S_{DBI} &=& -N_f \tau_p \int d^{p+1}y ~\frac{1}{2}  \sqrt{-det M_{ab}}~ M^{ab} \delta M_{ab}~,\\
\nonumber &=& -N_f \tau_p \int d^{p+1}y ~\frac{1}{2}  \sqrt{-det M_{ab}} \left(\ms^{ab}+\mathcal{A}^{ab}\right) \left(\delta \gamma_{ab}+ \delta F_{ab}\right)~,\\
&=& -N_f \tau_p \int d^{p+1}y ~\frac{1}{2}  \sqrt{-det M_{ab}} \left(\ms^{ab} \delta \gamma_{ab} + \mathcal{A}^{ab} \delta F_{ab}\right)~,
\eea
where $\ms^{ab}$ and $\mathcal{A}^{ab}$ are respectively the symmetric and anti-symmetric part of $M^{ab}$. So clearly
\be
\frac{\delta S_{DBI}}{\delta \gamma_{ab}}= - \frac{1}{2} N_f \tau_p  \sqrt{-det M_{ab}} \ms^{ab} ~.
\ee
Now to find $\ms$ and $\mathcal{A}$ we note that $M^{-1} M =1$. Taking transpose of both sides we get
\be
\gamma. \ms - F. \ms - \gamma .\ma + F. \ma = 1~,
\ee
where we have used the fact that $\gamma$ and $\ms$ are symmetric matrices whereas $F$ and $\ma$ are antisymmetric ones. Thus we have
\bea
\ms + \ma &=& \left(\gamma + F\right)^{-1}~,\\
\ms-\ma &=& \left(\gamma - F\right)^{-1}~.
\eea
On solving, we get 
\bea
\ms^{ab}&=& \left[\left(\gamma + F\right)^{-1} . \gamma .\left(\gamma -F\right)^{-1}\right]^{ab}~,\\
\ma^{ab} &=& -  \left[\left(\gamma + F\right)^{-1} . F.\left(\gamma - F\right)^{-1}\right]^{ab}~.
\eea
The open string metric is given by the inverse of $\ms^{ab}$,
\be
\ms_{ab} = \gamma_{ab} - \left(F. \gamma^{-1}.F\right)_{ab}~,
\ee
and its determinant is 
\be
det \ms =  det \gamma - \left(det F \right)^2 \left(det \gamma \right)^{-1}= \frac{det(\gamma+ F)~ det(\gamma - F)}{det \gamma }= \frac {(det M)^2}{det \gamma}~,
\ee
where in the last step we have used the fact that $ det(\gamma+ F)= det(\gamma- F)$. So we have
\be
-det M = \sqrt{det \gamma ~ det S}~,
\ee
which we have used in going from (\ref{bdy}) to (\ref{bdy2}).
\section{On Embedding function theorem:} \label{apenB}
In \cite{PhysRevD.100.126004},  the authors showed that for perturbative changes in area for spherical subregion, a general statement can be proved that restricts the order upto which, change in embedding function can contribute to a particular order change in area (HEE). The statement goes like following,
\begin{enumerate}
    \item Uncharged BH: $\Delta S^{(n)}$ ($n$-th order change in HEE) is determined by the embedding function up to and including $z_{\frac{n}{2}} (x)$ if $n$ is even or upto $z_{\frac{(n-1)}{2}}(x)$ if $n$ is odd.
    
    \item Charged BH: $\Delta S^{( \Vec{n} )}$ is determined by the embedding function up to and including $z_{\Vec{m}}(x)$, where $\Vec{m}$ is the highest possible order such that $| \Vec{m} | \leq \frac{| \Vec{n} |}{2}$.
\end{enumerate}
It was worth checking whether this theorem (uncharged case) holds in open string geometries as well and as it turns out, it does in all the dimensions we have considered. Although, we have not given the results for the third order changes in HEE, since they are physically not that important, but we do find that even for these non-Einstein solutions, the embedding function theorem holds and only first order change in embedding function is enough to produce the right third order change in area.

Just for an illustration, consider the $AdS_{5}$-osm case with spherical subregion where we have a fractional $3/2$-th order change in area (for HEE, we again write it in integer orders in terms of the electric field (\ref{hee5})). %due to the way we have defined the perturbation parameter, which is somewhat similar to the charged BH case studied in \cite{PhysRevD.100.126004}. %In this case also, we find that the first order change in embedding doesn't contribute to $3/2$-th order change in area (only the pure $AdS$ part of the embedding does) and the $3/2$-th order embedding doesn't contribute to the second order change in area. For convenience of the reader, we write%
The embedding functions upto $3/2$-th order turns out to be

\begin{equation}
    z^{(1)}(y) = \frac{1}{30} \left(1-y^2\right)^{3/2} \left(3 \left(y^2-2\right) \cos ^2(\theta ) \sin ^2(\phi )+5\right),
\end{equation}
\begin{equation}
    z^{(3/2)}(y) = \frac{1}{70} \left(1-y^2\right)^{3/2} \left(\left(5 y^4-13 y^2+11\right) \cos ^2(\theta ) \sin ^2(\phi )-5 \left(y^4-4 y^2+5\right)\right).
\end{equation}
It can be easily checked that $z^{(1)}(y)$ does not contribute to $\Delta \mathcal{A}^{(1)}$ or $\Delta \mathcal{A}^{(\frac{3}{2})}$ (expected as $\frac{3}{4}<1$). It only contributes to $\Delta \mathcal{A}^{(2)}$. Similarly $z^{(3/2)}(y)$ only starts contributing from $\Delta \mathcal{A}^{(n)}$, where $n \geq 3$. This validates the theorem.
\bibliographystyle{JHEP}
\bibliography{osmHEE}
\end{document}